\documentclass[preprint,3p,12pt]{elsarticle}
\usepackage{amssymb,amsmath,latexsym,braket,fancyref,hyphenat}

\numberwithin{equation}{section}

\biboptions{numbers,sort&compress}

\begin{document}

\begin{flushright}
CERN-PH-TH/2016-054\\ 
MAN/HEP/2016/05\\
March 2016\\[-1cm]
${}$
\end{flushright}

\vspace{3em}

\begin{frontmatter}

\title{
{\bf Frame-Covariant Formulation of Inflation\\
in Scalar-Curvature Theories}
}

\author[man]{Daniel Burns}
\ead{daniel.burns@manchester.ac.uk}
\author[man]{Sotirios Karamitsos}
\ead{sotirios.karamitsos@postgrad.manchester.ac.uk}
\author[man,cern]{and Apostolos Pilaftsis}
\ead{apostolos.pilaftsis@manchester.ac.uk}

\address[man]{Consortium for Fundamental Physics, School of Physics and Astronomy, University of Manchester, Manchester M13 9PL, United Kingdom}
\address[cern]{Theory Division, CERN, CH-1211 Geneva 23, Switzerland}

\begin{abstract} 
{\small  We develop a frame-covariant formulation of inflation in the slow-roll approximation by generalizing the inflationary attractor solution for scalar-curvature theories.  Our formulation gives rise to new generalized forms for the potential slow-roll parameters, which enable us to examine the effect of conformal transformations and inflaton reparameterizations in scalar-curvature theories. We find that cosmological observables, such as the power spectrum, the spectral indices and their runnings, can be expressed in a concise manner in terms of the generalized potential slow-roll parameters which depend on the scalar-curvature coupling function, the inflaton wavefunction, and the inflaton potential. We show how the cosmological observables of inflation are frame-invariant in this generalized potential slow-roll formalism, as long as the end-of-inflation condition is appropriately extended to become frame-invariant as well. We then apply our formalism to specific scenarios, such as the induced gravity inflation, Higgs inflation and $F(R)$ models of inflation, and obtain more accurate results, without making additional approximations to the potential. Our results are shown to be consistent to lowest order with those presented in the literature.  Finally, we outline how our frame-covariant formalism can be naturally extended beyond the tree-level approximation, within the framework of the Vilkovisky--DeWitt effective action. 

}
\end{abstract}


\end{frontmatter}

\section{Introduction}
\label{sec:intro}

Inflation, which was originally proposed as a solution to the flatness and horizon problems \cite{Guth:1980zm,Linde:1981mu}, has been found to be an excellent generic explanation to the origin of anisotropies observed in the cosmic microwave background (CMB) \cite{Hawking:1982cz,Starobinsky:1982ee,Guth:1982ec,Linde:1983gd}.  However, the large number of inflationary models underpinned by a variety of theoretical ideas, such as quintessence, modified gravity and string theory, poses a challenge in determining the physical driving mechanism for inflation. Furthermore, due to the complexity of the equations of motion in many inflationary models, extracting analytical predictions for cosmological observables of inflation is rapidly becoming a formidable task. In the simplest of inflationary models, it has been found that if the equations of motion for the classical perturbations of the metric and the inflaton are quantized, the observed tilt of the CMB can be found in terms of solutions to the classical equations of motion. To analytically investigate inflationary models in the general case, the standard procedure is to assume that these solutions satisfy a set of constraints known collectively as the \emph{slow-roll approximation}. In this paradigm, the inflaton field $\varphi$ is assumed to slowly roll down the inflationary potential~$V(\varphi )$, meaning that we may neglect several terms in the equations of motion. Consequently, it is possible to obtain simple analytical expressions for the tilt of the CMB and other inflationary observables.

As observations impose increasingly tighter constraints on inflation \cite{Ade:2015lrj,Ade:2015tva}, it becomes difficult to physically motivate minimally coupled inflationary models with acceptable phenomenology both in the context of particle physics and cosmology. A popular alternative is to introduce a coupling function $f(\varphi )$ between the scalar curvature~$R$ and the inflaton $\varphi$, leading to a more general class of gravity models, termed \emph{scalar-curvature theories}. In these theories, such a coupling function may also be motivated by viewing it as emerging from quantum corrections to the low-energy effective action, after integrating out high-energy degrees of freedom. Thus, it is desirable to extend the procedure for extracting observable quantities from minimally coupled models, in which $f(\varphi ) = M^2_P$ where $M_P = 2.435 \times 10^{18}~{\rm GeV}$ is the reduced Planck mass, to general scalar-curvature models, in which $f(\varphi )$ is an arbitrary function of $\varphi$. Moreover, scalar-curvature theories with a non-trivial scalar-curvature coupling $f(\varphi )$, which are said to be in the \emph{Jordan frame}, can be recast in the \emph{Einstein frame}, and so be written in terms of minimally coupled models via a combination of conformal transformations and inflaton field reparameterizations. Consequently, studying models related by these transformations can help resolve the so-called \emph{frame problem}, namely whether these models are physically equivalent or not~\cite{Dicke:1961gz,Gasperini:1993hu,faraoni98,Steinwachs:2013tr,Chiba:2013mha, Postma:2014vaa,Kamenshchik:2014waa,Domenech:2015qoa}.

The aim of the present article is to introduce frame covariance in the  inflationary dynamics of scalar-curvature theories. This covariance manifests itself as a set of transformation rules that nonetheless keep cosmological observables of inflation invariant. To this end, we develop a new formalism for extracting predictions for observable quantities from scalar-curvature theories by generalizing the corresponding well-known potential slow-roll approximation used in minimally-coupled models. Using this formalism, it is possible to study classes of scalar-curvature theories related to one another by conformal transformations and inflaton field reparameterizations independently. Hence, we will show that these classes of models generate equivalent predictions for inflationary observables. Furthermore, the new formalism may be used as a calculational tool for extracting predictions in a concise and intuitive manner for a wide range of scalar-curvature models without the need for further approximations beyond the ones established in the  slow-roll approximation.

The outline of this paper is as follows: after this introductory section, Section \ref{sec:modspec} introduces the classical action $S$ of the scalar-curvature theories that we will be considering. In particular, we specify each theory by three {\it model functions}~[cf.~\eqref{actionJ}]: (i)~the non-minimal scalar-curvature coupling~$f(\varphi )$, (ii)~the non-canonical inflaton-dependent wavefunction~$k(\varphi )$, and (iii)~the inflaton potential~$V(\varphi )$. We introduce conformal transformations and inflaton reparameterizations and, by observing that the classical action of the theory is invariant under their combined action, we derive the transformation properties of the three model functions mentioned above.

In Section \ref{sec:classdyn}, we derive the modified Einstein field equations for scalar-curvature theories by varying the action with respect to the metric~$g_{\mu\nu}$ and the inflaton field~$\varphi$. We further simplify the cosmological equations of motion by considering a Friedman--Robertson--Walker~(FRW) metric with a general lapse function~$N_L$ and a homogeneous inflaton. We observe that the form of the acceleration, Friedman, and continuity equations does not alter, as long as the energy density and pressure are replaced by new, modified variables given in terms of the model functions.

In Section \ref{sec:cosmpert}, we perturb the metric to first order, which allows us to consider separately scalar, vector, and tensor perturbations. We then introduce the comoving curvature perturbation as the primordial origin of scalar perturbation modes and the polarizations of the gravitational waves as the primordial tensor perturbation modes.  After quantizing these perturbations, we write down their two-point correlation functions and relate the latter to the scalar and tensor power spectra~$P_{\cal R}$ and~$P_T$. In this way, we introduce the commonly used inflationary observables in terms of~$P_{\cal R}$ and $P_T$, which include the scalar and tensor spectral indices $n_{\cal R}$ and $n_T$, the tensor-to-scalar ratio~$r$, and the runnings of the spectral indices~$\alpha_{\cal R}$ and $\alpha_T$.

In Section \ref{sec:srinf}, we introduce the slow-roll approximation by defining the Hubble slow-roll parameters, which allow us to neglect certain terms in the equations of motion and control the validity of the approximation. However, the presence of a non-trivial non-minimal coupling $f(\varphi )$ is found to introduce two extra slow-roll parameters in addition to those present in minimally-coupled models of inflation. After writing the inflationary observables mentioned above and the cosmological equations of motion in terms of the Hubble slow-roll parameters, we derive the generalized solution for the inflationary attractor. This enables us to define new {\em potential slow-roll parameters}, purely in terms of~$f(\varphi )$, $k(\varphi )$ and~$V(\varphi )$, which reduce to the Hubble slow-roll parameters in the slow-roll approximation. Hence, we derive explicit expressions for the inflationary observables in a straightforward manner for any scalar-curvature theory using only the expressions of the model functions, thus avoiding the intermediate step of having to solve the equations of motion.

In Section \ref{sec:reparam}, we examine the different frames that may occur in inflationary dynamics and derive the transformation properties of the generalized potential slow-roll parameters. By virtue of these parameters, we show that cosmological observables of inflation are frame-independent when expressed in terms of the inflaton~$\varphi$ to first order in the slow-roll approximation. Instead, the number of e-folds, commonly used in the literature to express analytic predictions for inflationary observables, is found to be frame-{\em dependent}. However, in our generalized approach, the end-of-inflation condition may be uniquely extended, so that it becomes frame-independent and reduces to the usual condition for the Einstein frame, thus leaving observables expressed in terms of e-folds frame-invariant.

In Section \ref{sec:specmod}, we consider three specific models of inflation: (i)~induced gravity inflation, (ii)~Higgs inflation, and (iii)~$F(R)$ theories. In induced gravity inflation, the effective Planck mass is fully induced by the inflaton. We distinguish between small-field and large-field induced gravity inflation and derive expressions for the cosmological observables in both cases.  We then proceed similarly in Higgs inflation, which contains a non-minimal coupling that modifies, but not fully dominates the effective Planck mass. By analogy, we derive expressions for all cosmological observables of inflation and evaluate the size of the non-minimal coupling through the normalization of the power spectrum. In all cases, the expressions for the cosmological observables reduce to the ones found in the literature to lowest order. Finally, we consider a slightly different class of theories, the so-called $F(R)$ models, where inflation is driven by a modification to the Einstein--Hilbert action. Using an auxiliary field, we recast these models in terms of scalar-curvature models, and so obtain predictions for a Starobinsky-like model, for which $F(R) = \alpha R + \beta_n R^n$.

In Section~\ref{sec:BCA}, we outline how our frame-covariant formalism can be extended beyond the tree-level approximation, within the framework of the Vilkovisky--DeWitt effective action~\cite{Vilkovisky:1984st,DeWitt}. Our explicit demonstration will be at the one-loop level, thus making plausible its applicability to higher orders.  Finally, Section \ref{sec:disc} summarizes our conclusions and presents possible future directions along the frame-covariant formalism for inflation that we are studying.  Technical details related to the transformation properties of the model functions are given in~\ref{appendix}.

\section{Scalar-Curvature Theories and Frame Transformations}
\label{sec:modspec}

In this section, we define the classical action $S$ describing the inflationary dynamics
in scalar-curvature theories. The invariance of $S$ under conformal rescalings of the metric $g_{\mu\nu}$ and repara\-meterizations of the inflaton field~$\varphi$ will help us to introduce the concept of {\it frame transformations}.

For simplicity, let us assume that the energy densities of all other fields are sufficiently diluted with respect to the energy density of the inflaton, such that there is no contribution to the Lagrangian from hydrodynamic matter. With this assumption, we may define the classical action $S$ for a wide class of scalar-curvature theories as
\begin{equation}
   \label{actionJ}
S[g_{\mu\nu}, \varphi, f(\varphi), k(\varphi),V(\varphi)]\ \equiv\   \int  d^4 x\,  \sqrt{-g}  \, \left[\, -\,\frac{f(\varphi)}{2} R\: +\: \frac{k(\varphi)}{2} \, g^{\mu\nu }(\nabla_\mu \varphi) (\nabla_\nu \varphi)\: -\: V(\varphi)\, \right]\;,
\end{equation}
where  $g \equiv \det g_{\mu\nu}$  and $R$ is the Ricci scalar. In addition, 
$f(\varphi )$ is the non-minimal scalar-curvature coupling function, $k(\varphi )$ is the non-canonical inflaton wavefunction, and $V (\varphi )$ is the inflaton potential. 
We collectively call the functions $f(\varphi )$, $k(\varphi )$ and $V (\varphi )$ 
{\it model functions} that enter the action $S$ in~(\ref{actionJ}).
We will adopt the convention $\eta_{\mu\nu} = {\rm diag}(+1,-1,-1,-1)$ for the Minkowski flat limit of $g_{\mu\nu}$,  and work in natural units where the mass parameters are normalized to the reduced Planck mass $M_P \equiv (8 \pi G)^{-1/2}$. Finally, we
define the covariant derivatives denoted by $\nabla_\mu$ to be metric-compatible with respect to $g_{\mu\nu}$, meaning that the action is diffeomorphism invariant.

By specifying the model functions $f(\varphi)$, $k(\varphi)$, and $V(\varphi)$, we can cover a wide range of models. For instance, the so-called $F(R)$ theories may be described by setting $k(\varphi) = 0$. More details are given in Section~\ref{FRmod}.
In all scenarios, we assume that the inflaton relaxes in its expected value $\varphi_\text{VEV}$ at the end of inflation, and so the effective reduced Planck mass $M_P$ matches its observed value at the present epoch, i.e.~$f(\varphi_\text{VEV}) = M_P^2 \equiv 1$. For a review of the dynamics 
of minimal inflationary scenarios, the reader may consult~\cite{Lyth:1998xn}.

It is now important to study the response of the classical action $S$ under conformal rescalings of the metric $g_{\mu\nu}$ and reparameterizations of the inflaton field~$\varphi$.  
To this end, we first perform a \emph{conformal transformation} by rescaling the metric 
\begin{equation}
   \label{weyl} 
g_{\mu\nu}\ \rightarrow\ \: \tilde g_{\mu\nu}\: =\: \Omega^2\, g_{\mu\nu}\;,
\end{equation} 
where the coordinate-dependent function $\Omega = \Omega (x)$ is known as the \emph{conformal factor}. Changing its value is often referred to as changing the \emph{conformal frame} of the theory. Under the conformal transformation \eqref{weyl}, the Ricci scalar transforms as 
\begin{equation}
   \label{riccitransform} 
{\widetilde R}\ =\ \Omega^{-2}R\: -\: 6\,\Omega^{-3} g^{\mu\nu}\nabla_\mu \nabla_\nu\Omega\;.  
\end{equation} 
Likewise, we may perform an arbitrary inflaton reparameterization $\varphi \to \tilde \varphi = \tilde \varphi (\varphi )$, whose explicit form may be determined by 
\begin{equation}
  \label{Kphi} 
  \left( \frac{d \tilde \varphi}{d\varphi} \right)^2\ =\ K (\varphi)\; .  
\end{equation}
Then, using \eqref{weyl} and \eqref{riccitransform} in \eqref{actionJ}, 
the classical action $S$, upon neglecting a total derivative, can be rewritten in the form
\begin{equation}
   \label{actionI}
 S[\tilde g_{\mu\nu}, \tilde\varphi, \tilde f( \tilde \varphi), \tilde k( \tilde\varphi),\widetilde V(\tilde  \varphi)]\ =\
 \int d^4 x \, \sqrt{ - \tilde g}\,  
 \left[ \,    
   -\, \frac{  \tilde f( \tilde  \varphi)}{2}  {\widetilde R}\: 
   +\:  \frac{\tilde k( \tilde \varphi)}{2} \, \tilde g^{\mu\nu} 
   (\nabla_\mu \tilde\varphi)  (\nabla_\nu \tilde\varphi )\:
   -\:  { \widetilde V(  \tilde \varphi)} 
 \right]\; .
\end{equation}
In the above, the transformed model functions $\tilde f(\tilde \varphi)$, $\tilde k( \tilde \varphi)$ and $\widetilde V(  \tilde \varphi)$ have been expressed in terms of the original ones $f(\varphi)$, $k(\varphi)$ and $V(\varphi)$ as follows~\cite{Flanagan:2004bz,Jarv:2014hma}:
\begin{eqnarray}
   \label{Imodelparamdef}
\tilde f (  \tilde \varphi) &=& \Omega^{-2}\, f\; , \nonumber \\
\tilde k( \tilde \varphi) &=& \frac{\Omega^{-2}}{K}\, \Big( k \:
-\: 6\, f\,\Omega^{-2}\Omega_{,\varphi}^2\:
+\:  6\, \Omega^{-1}    f_{,\varphi}\, \Omega_{,\varphi}  \Big)\; ,\\
\widetilde V(  \tilde\varphi) &=& \Omega^{-4}\, V\; .\nonumber
\end{eqnarray}
Here, $f(\varphi)$, $k(\varphi)$ and $V(\varphi)$, the conformal factor $\Omega$, and their possible derivatives with respect to~$\varphi$, appearing on the right-hand side (RHS) of~(\ref{Imodelparamdef}), all depend on~$\varphi$.  Alternatively, they may also be expressed in terms of the transformed field $\tilde \varphi$ through $\varphi = \varphi(\tilde \varphi)$, after inverting the solution $\tilde \varphi = \tilde \varphi (\varphi )$ to Equation~\eqref{Kphi}. Technical details related to the derivation of the transformation properties of the model functions given in~(\ref{Imodelparamdef}) are presented in~\ref{appendix}.

The original action \eqref{actionJ} is said to be in the {\it Jordan frame}, where the non-minimal coupling of the inflaton to the curvature appears explicitly. In most analyses of inflationary dynamics, one usually considers the {\it Einstein frame}, for which the conformal factor $\Omega$ is chosen such that the non-minimal coupling becomes minimal: $\tilde f(\tilde \varphi) = M_P^2$. However, in this article, we will be more general and consider the full class of conformal transformations, where $\Omega = \Omega (x)$ is an arbitrary well-behaved function. {\em While frame invariance is often assumed as an `a priori' principle, there is no guarantee that any given theory will generate frame-independent predictions unless it has explicitly been constructed to be frame-invariant.} As such, there has been much discussion about whether the Jordan frame or the Einstein frame are physically equivalent.  In particular, there have been claims of both conformal independence \cite{Chiba:2013mha} and conformal dependence \cite{vollick03} in the literature. This is further compounded by the fact that after a conformal tranformation, the wavefunction $k(\varphi)$ of the inflaton kinetic term is not canonical. Hence, some authors include in their definition of ``conformal transformations'' a field reparameterization that renders the kinetic term canonical, i.e.~$k(\varphi) \to 1$. For this reason, we shall use a more general terminology and call the combined effect of a conformal transformation and a field reparameterization a {\it frame transformation}.

From the above discussion, it has become clear that the functional form of the classical action~$S$ as defined in~(\ref{actionJ}) remains invariant under general frame transformations [cf.~(\ref{actionI})]. The functional form of~$S$ could have been modified, for example, by the presence of higher-order derivative terms induced by the conformal rescaling~\eqref{weyl}. This means that under frame transformations, the action of a scalar-curvature theory  gets transformed to an equivalent action within the same class of theories. This basic property of invariance of the classical action $S$ in~\eqref{actionJ} under frame transformations may be expressed as follows: 
\begin{equation}
   \label{actiontransprop}
S[g_{\mu\nu}, \varphi, f(\varphi), k(\varphi),V(\varphi)]\  =\ S[\tilde g_{\mu\nu}, \tilde \varphi, \tilde f(\tilde \varphi), \tilde k(\tilde \varphi),\widetilde V(\tilde \varphi)]\; .
\end{equation}
Note that although the functional form of $S$ does not change, the functions $f$, $k$ and $V$ {\em do} change as given in~\eqref{Imodelparamdef}, as a consequence of frame transformations. Equation~\eqref{actiontransprop} represents a fundamental property that underlies our frame-covariant formulation of inflation. In Section~\ref{sec:BCA}, we will show how this fundamental property~\eqref{actiontransprop} can be extended to the effective action beyond the tree-level approximation. Thus, developing a formalism that can be applied to a general scalar-curvature theory will allow us to independently examine and compare the predictions for the inflationary observables that are obtained by using $S[g_{\mu\nu}, \varphi, f(\varphi), k(\varphi),V(\varphi)]$ {\em or} $S[\tilde g_{\mu\nu}, \tilde \varphi, \tilde f(\tilde \varphi), \tilde k(\tilde \varphi),\widetilde V(\tilde \varphi)]$.  This exercise will be useful to address the question of whether  frame transformations are physically significant or not. Our first step towards developing such a formalism will be to study the behaviour of the background fields during inflation in the next section.

\section{Classical Dynamics}
\label{sec:classdyn}

In this section, we  consider the cosmological evolution of the background inflaton field $\varphi$, since its imprint on observable quantities depends on the value of $\varphi$ at horizon exit. Under the assumption that the inflaton $\varphi$ is spatially homogeneous evolving in a space described by the FRW metric, we~derive the equations of motion for $\varphi$ by treating it as a perfect fluid.

Taking now the functional derivative of the action \eqref{actionJ} with respect to $\varphi$ yields the inflaton equation of motion
\begin{equation}
   \label{infeq}
k\,\nabla^2 \varphi\: +\: \frac{k_{,\varphi}}{2}\,(\nabla  \varphi)^2\: +\: V_{,\varphi}\: 
+\: \frac{f_{,\varphi} }{2}\, R\ =\ 0\; ,
\end{equation}
where $_{,\varphi}$ denotes differentiation with respect to $\varphi$ and we suppress arguments of $\varphi$ from now on. Similarly, by varying \eqref{actionJ} with respect to the metric~$g_{\mu\nu}$, we obtain the generalized Einstein equation
\begin{eqnarray}
   \label{eomg}
G_{\mu\nu}\ \equiv\   R_{\mu\nu}\: -\: \frac{1}{2}g_{\mu\nu}R 
&=&
\frac{T_{\mu \nu}}{f}\: -\: \frac{f_{,\varphi\varphi}}{f}\, (\nabla  \varphi)^2\, g_{\mu\nu} \:
-\: \frac{f_{,\varphi}}{f}\,(\nabla^2 \varphi) \, g_{\mu\nu}  
\nonumber\\
&&
+\: \frac{f_{,\varphi}}{f}(\nabla_\mu \nabla_\nu \varphi) \:  
+\: \frac{f_{,\varphi\varphi}}{f} (\nabla_\mu \varphi) ( \nabla_\nu \varphi)\; ,
\end{eqnarray}
where $G_{\mu\nu}$ is the Einstein tensor, $R_{\mu\nu}$ is the Ricci tensor and $T_{\mu\nu}$  is the energy-momentum tensor  defined as
\begin{equation}
   \label{set}
T_{\mu\nu}\ \equiv\ \frac{2}{\sqrt{-g}}\,\frac{ \delta S_\varphi}{\delta g^{\mu\nu}}\; ,
\end{equation}
where $S_\varphi$ is the matter part of the inflaton action.
Applying this last formula to the matter sector given in~\eqref{actionJ}, we find
\begin{equation}
   \label{setphi}
T_{\mu\nu}\  =\ 
 k\,  (\nabla_\mu \varphi) (\nabla_\nu \varphi) \:
-\: \frac{k }{2}\,(\nabla \varphi)^2 g_{\mu\nu}\: 
+\: V g_{\mu\nu} \; .
\end{equation}
Contracting \eqref{eomg} with the inverse metric~$g^{\mu\nu}$, we can eliminate $R$ from \eqref{infeq}, leading to a more convenient form for the inflaton equation,
\begin{equation}
   \label{eomg2}
\bigg( k\:  +\: \frac{3\,f_{,\varphi} ^2}{2\,f} \bigg)\,\nabla^2 \varphi \
+\ \bigg( 
\frac{k_{,\varphi}}{2}\:
+\:  \frac{3\,f_{,\varphi} }{2\,f} f_{,\varphi\varphi}\:
+\: \frac{ k}{2}\, \frac{f_{,\varphi} }{f}  \bigg) (\nabla  \varphi)^2 \
+\ f^2 U_{,\varphi}\ =\ 0\; ,
\end{equation}
with
\begin{equation}
   \label{Udef}
 U\ \equiv\ \frac{V}{f^2}\ .
\end{equation}
We observe  that \eqref{eomg} may be written in the standard form for the Einstein equation
as
\begin{equation}
   \label{einmod}
 R_{\mu\nu}\: -\: \frac{1}{2}\,g_{\mu\nu}\,R\ = \  M_P^{-2}\,T^{\rm(NM)}_{\mu\nu}\; .
\end{equation}
Here, $T^{\rm(NM)}_{\mu\nu}$ is the modified, non-minimal (NM) energy-momentum tensor defined as
\begin{eqnarray}
   \label{effset}
\frac{T^{\, \rm (NM)}_{\mu\nu}}{M_P^2}   &\equiv& 
\frac{T_{\mu \nu}}{f}\ -\ \frac{f_{,\varphi\varphi}}{f}\,(\nabla  \varphi)^2\,g_{\mu\nu}\ 
-\ \frac{f_{,\varphi}}{f}\,(\nabla^2 \varphi) \, g_{\mu\nu}  
\nonumber\\
& &
+\: \frac{f_{,\varphi}}{f}\,(\nabla_\mu \nabla_\nu \varphi)\  
+\ \frac{f_{,\varphi\varphi}}{f}\,(\nabla_\mu \varphi) ( \nabla_\nu \varphi)\; .
\end{eqnarray}
Evidently, the standard Einstein gravity is recovered when $f=M_P^2$ and $T^{\, \rm (NM)}_{\mu\nu}$ is replaced with~$T_{\mu\nu}$. 

The equations of motion of cosmological interest can be derived under the assumption that $\varphi = \varphi(t)$ is spatially homogeneous and that the universe is described by a flat FRW metric of the form $g_{\mu\nu} = \text{diag}(N_L^2,-a^2,-a^2 ,-a^2)$, where $a = a(t)$ is the scale factor and $N_L = N_L(t)$ is the lapse function. Imposing these conditions on the modified energy-momentum tensor $T^{\, \rm (NM)}_{\mu\nu}$, we find that $T^{\, \rm (NM)}_{\mu\nu}$ is diagonal, which enables us to define the modified energy density $\rho^{\, \rm (NM)}$ and pressure~$\, p^{\, \rm  (NM)}$~as \cite{Akbar:2006er}
\begin{equation}
   \label{effrhop}
T^{\, \rm (NM)}_{\mu\nu}\ \equiv\ \text{diag}\left(N_L^2 \rho^{\, \rm (NM)}, -\,a^2 p^{\, \rm (NM)},  -\, a^2 p^{\, \rm (NM)} , -\,a^2 p^{\, \rm (NM)} \right)\; .
\end{equation}
In this way,  using \eqref{effrhop}, the explicit forms of the modified energy density and pressure are found to be
\begin{eqnarray}
    \label{rhoeff}
\frac{\rho^{\, \rm  (NM) }}{M_P^2}
&=&
 \frac{\rho}{ f}\ -\  \frac{3H {\dot f}}{ f}  ,\\
    \label{peff}
\frac{p^{\, \rm  (NM)}}{M_P^2}
&=&
\frac{p}{f}\ 
+\    \frac{2H {\dot f}}{ f}\
+\ \frac{{\ddot f}}{ f}\ .
\end{eqnarray}
Here and in the following, the Hubble parameter is defined as
\begin{align}
   \label{Hubble}
 H\ \equiv\ \frac{\dot a}{a}\ ,
\end{align}
where the overdot from now on denotes differentiation with respect to $\tau$. The latter is related to the cosmic time~$t$ through $d\tau \equiv  N_L dt$ and includes the effect of the general lapse function $N_L$. Thus, $\dot a$ is defined, for instance, as
 \begin{align}
 \dot a\ \equiv\ \frac{1}{N_L}\,\frac{da}{dt}\ .
\end{align}
In addition, $\rho$ and $p$ denote the ordinary comoving energy density and pressure, respectively, as these are read off from $T_{\mu\nu}\equiv \text{diag}(N_L^2\rho,-a^2 p,-a^2 p,-a^2 p)$, i.e.
\begin{equation}
\rho\ =\ \frac{k}{2}\, \dot \varphi^2\: +\: V\;,\qquad
p\ =\ \frac{k}{2}\, \dot \varphi^2\: -\:  V\; .
\end{equation}
With the definitions of $\rho^{\rm (NM)}$ and  $p^{\rm(NM)}$ given in \eqref{rhoeff} and \eqref{peff}, the continuity, Friedmann, and the acceleration equations take on the form
\begin{eqnarray}
   \label{contNM}
&&\hspace{-1.2cm}\dot \rho^{\, \rm  (NM)}\: +\: 3 H \left[\rho^{\, \rm  (NM)}\: +\: p^{\, \rm (NM)}\right]\  =\ 0\; ,\\
   \label{friedNM}
H^2  &=&
  \frac{\rho^{\,  \rm (NM)}}{3 }\ ,\\
   \label{accelNM}
\dot H &=& - \frac{\rho^{\, \rm  (NM)}+p^{\, \rm  (NM)}}{2}\ .
\end{eqnarray} 
These equations become identical to the minimal case, for $\rho^{\rm (NM)}\to \rho$ and $p^{\rm(NM)}\to p$. Substituting the forms of $\rho^{\rm (NM)}$ and $p^{\rm(NM)}$ given in \eqref{rhoeff} into \eqref{contNM}, \eqref{friedNM} and \eqref{accelNM}, we derive cosmological equations of motion for the general scalar-curvature theories~\cite{Kaiser:1994vs}, 
\begin{eqnarray}
   \label{eomj1}
&&\hspace{-1.2cm}
\bigg( k\:  +\: \frac{{3f ^2_{,\varphi}}}{2f} \bigg)\, 
\Big({\ddot \varphi} + 3H{\dot \varphi}\Big)\
+\ \bigg( \frac{k_{,\varphi}}{2}\: +\: \frac{3f_{,\varphi} }{2f}\,f_{,\varphi\varphi}
+ \frac{ k}{2}\,\frac{f_{,\varphi} }{f}  \bigg) {\dot \varphi}^2\
+\ f^2 U_{,\varphi}\ =\ 0\; ,\\
   \label{eomj2}
H^2  &=&  \frac{1}{3f}\bigg( \frac{k}{2}\, {\dot \varphi}^2\:  +\: V\bigg)\ -\ 
\frac{f_{,\varphi}}{f}  H\,{\dot \varphi}\; ,\\
   \label{eomj3}
\dot H &=& - \frac{k \dot \varphi^2  }{2f}\ +\  \frac{ H\dot f  }{2 f}\ - \ \frac{  \ddot f}{2f}\ .
\end{eqnarray}
Notice that these equations are written down by neglecting the spatial dependence of the background inflaton field~$\varphi$, i.e.~$\varphi = \varphi(t)$.

The cosmological equations of motion that we have presented here for the background metric~$g_{\mu\nu}$ and the inflaton field~$\varphi$ will be useful for our discussions in the subsequent sections. Specifically, the general equations of motion for~$\varphi$ and~$g_{\mu\nu}$, stated in~\eqref{eomg} and \eqref{eomg2}, will be needed to study the cosmological perturbations in Section~\ref{sec:cosmpert}. Likewise, the homogeneous limit of these equations presented in \eqref{eomj1}, \eqref{eomj2} and~\eqref{eomj3} will be used to determine the inflationary attractor trajectory for the scalar-curvature theories. The latter will in turn be used in Section~\ref{sec:srinf} to evaluate the predictions for cosmological observables of inflation in these theories.

\section{Cosmological Perturbations}
\label{sec:cosmpert}

The imprint left by inflation on the CMB can be understood in terms of perturbations to the metric and the inflaton. At very early times, perturbations corresponding to scales of cosmological interest (smaller than the current size of the observable universe) are inside the Hubble horizon and are free to evolve \cite{Weinberg:2003sw}. When the perturbations leave the horizon, they stop evolving \cite{Weinberg:2008nf}, meaning that the observed anisotropy in the CMB at scales entering the horizon in the current epoch was formed at the point of horizon exit. This anisotropy can be found by calculating the two-point correlation function of cosmological perturbations just before they exited the horizon. In~this section, we will depart from the well-studied perturbations in minimally-coupled inflationary models \cite{Lyth:1998xn} and consider perturbations of scalar-curvature theories. In order to make contact with observations, we will look at how cosmological perturbations evolve with the aim of calculating the power spectrum of the CMB. We will start by writing the explicit form of the perturbation of the metric, which we will use to write down the linearized Einstein field equations to first order which control how the classical perturbations evolve, allowing us to quantize them and to calculate the two-point function, from which we may make contact with inflationary observables.

The first step in determining the evolution of the quantized perturbations is to study the evolution of classical perturbations. This is given by the linearized Einstein field equations, \begin{align}\label{lineqein} \delta G_{\mu\nu}\ =\ M_P^{-2}\,\delta T^{\, \rm (NM)}_{\mu\nu}, \end{align} where $\delta G_{\mu\nu}$ and $\delta T^{\, \rm (NM)}_{\mu\nu}$ are the perturbations of the respective Einstein and energy-momentum tensors, $G_{\mu\nu}$ and $T_{\mu\nu}$, that result from linear variations of $\varphi$ and $g_{\mu\nu}$. Explicitly, the inflaton~$\varphi$ and the metric~$g_{\mu\nu}$ may be expanded around their background values $\bar \varphi$ and $\bar g_{\mu\nu}$ as 
\begin{equation} 
\varphi\ =\ \bar \varphi\: +\: \delta \varphi\; , \qquad 
g_{\mu\nu}\ =\ \bar g_{\mu\nu}\: +\: \delta g_{\mu\nu}\; .  
\end{equation} 
Parameterizing the perturbation $\delta g_{\mu\nu}$ of the metric is more involved, as there are several degrees of freedom that need to be taken into account. Therefore, we adopt the standard scalar-vector-tensor decomposition and expand the full metric~$g_{\mu\nu}$ as follows:
\begin{eqnarray}
    \label{metricpar}
g_{\mu\nu} dx^\mu dx^\nu  &=&  (1+ 2\phi) N_L^2 dt^2\ +\ 2 a(\partial_i B + B_i) N_L   dt\,  dx^i
\nonumber \\
&&-\: a^2 \Big[ (1 +  2\psi) \delta_{ij} + \partial_i \partial_j A + \partial_i A_j + \partial_j A_i + h_{ij}\Big]\, dx^i dx^j\; ,
\end{eqnarray}
where $\phi, \psi, A, B$ are the scalar perturbations, $A_i, B_i$ are the vector perturbations and $h_{ij}$ is the tensor perturbation to the metric.  All these perturbations are independent of each other.

As the only persistent (scalar) measure of inflation, we introduce the diffeomorphism-invariant \emph{comoving curvature perturbation} $\mathcal{R}_\varphi$ \cite{Bardeen:1980kt}:
\begin{align}
\mathcal{R}_\varphi\ =\ \phi\: -\: \frac{H}{\dot {\bar \varphi}} \delta {  \varphi}\; ,
\end{align}
where $\phi$ is the remaining physical degree of freedom in the perturbed metric. After expanding the $00$, $0i$, and $ij$ components of the linearized Einstein equations, we find that the comoving curvature perturbation $\mathcal{R}$ satisfies the following equation in Fourier space
\cite{Hwang:1996xh}:
\begin{align}\label{smeq}
\frac{1}{N_L^2 a^3 Q_\mathcal{R}}\frac{d}{dt} \left( N_L a^3 Q_\mathcal{R}   
{\mathcal{\dot R}} \right)\ +\ \frac{k^2  \mathcal{R}}{a^2}\ =\ 0\;,
\end{align}
where $k \equiv |{\bf k}|$ corresponds to the scale of the Fourier mode $ {\bf k}$ of the perturbation, and $\mathcal{R}$ corresponds to the Fourier components of the comoving curvature perturbation. Similarly, the two polarizations of the gravitational waves $h_{+,\times}$ satisfy the following equation:
\begin{equation}
   \label{smeqT}
\frac{1}{N_L^2 a^3 Q_T}\frac{d}{dt} \left( N_L a^3 Q_T   {\dot h_{+,\times}} \right)\ +\ \frac{k^2  h_{+,\times}}{a^2}\  =\ 0\; .
\end{equation}
The quantities $Q_\mathcal{R}$ and $Q_T$ are given by
\begin{equation}
   \label{smeqdef}
Q_\mathcal{R}\ =\ \frac{k \dot \varphi^2 + \frac{3 \dot f^2}{2f} }{\left( H + \frac{\dot f}{2f}\right)^2 }\ \equiv\ \frac{\dot \varphi^2}{H^2} Z_\mathcal{R}\; ,
\qquad
Q_T\ =\ f \ \equiv\  M_P^2 Z_T\; ,
\end{equation}
where, for later convenience, we have defined $Z_\mathcal{R}$ and $Z_T$ as
\begin{equation}
   \label{smeqdefQ}
 Z_\mathcal{R}\ \equiv\
\frac{k + \frac{3 \dot f^2}{2f \dot \varphi^2 } }{\left( 1 + \frac{\dot f}{2Hf}\right)^2 }\ ,
\qquad
Z_T \  \equiv\ \frac{f}{M_P^2}\ .
\end{equation}
Note that $Z_\mathcal{R} = Z_T = 1$ in the Einstein frame, in which $f = M_P^2$ and $k = 1$. By further defining 
\begin{equation}
    \label{zdef}
z_\mathcal{R}\ \equiv\ a \sqrt{Q_\mathcal{R}}\; ,
\qquad
v_\mathcal{R}\ \equiv\ z_\mathcal{R} \mathcal{R}\; ,
\end{equation}
and similarly for $z_T$ and $v_T$,
\begin{equation}
   \label{zdef2}
z_T\ \equiv\ a \sqrt{Q_T}\; ,
\qquad
v_T\ \equiv\  z_T  h_{+,\times}\; ,
\end{equation}
the equations of motion \eqref{smeq} and \eqref{smeqT} can be written as
\begin{eqnarray}
    \label{smstan}
\frac{d^2 v_{\mathcal{R},k}}{d\eta^2}\  +\ \left(k^2 -\frac{1}{z_\mathcal{R}}\frac{d^2 z_\mathcal{R}}{d\eta^2}  \right) v_{\mathcal{R},k} &=& 0\; ,\\
\frac{d^2 v_{T,k}}{d\eta^2}\ +\ \left(k^2 -\frac{1}{z_T}\frac{d^2 z_T}{d\eta^2}  \right) v_{T,k} 
&=& 0\; ,
\end{eqnarray}
which correspond to simple harmonic oscillators with time-dependent masses and the conformal time $\eta$ is given by $N_L dt = a d\eta$. Treating $v_\mathcal{R}$ and its conjugate momentum as operators and imposing the usual commutation relations on them, we can write down its mode expansions in terms of the mode functions $v_{\mathcal{R},k}$ and the corresponding creation and annihilation operators. Imposing the condition that, in the early time limit, the perturbations live in \emph{de Sitter space}, which is characterized by a constant vacuum density driving its accelerated expansion and $\eta = -1/(aH)$, we may write the mode functions that satisfy \eqref{smstan}. Then, from the condition that the solutions must correspond to the Bunch-Davies vacuum at very early times \cite{Allen:1985ux}, we can finally write the two-point correlation function for the canonical fields $v_\mathcal{R}$,
\begin{equation} 
\braket{v_{\mathcal{R}, {\bf k}_1}\vert v_{\mathcal{R}, {\bf k}_2}}\ =\ 
\left\vert v_{\mathcal{R},k} \right\vert^2\, \delta({\bf k}_1 + {\bf k}_2)\; ,
\end{equation}
This is the correlation function for fields in de Sitter space, but since inflation ends when the comoving horizon stops shrinking, the fields are in \emph{quasi-de Sitter} space, and their two-point correlation function is related to that of the canonical fields by the 
normalization~\eqref{zdef}. For scalar perturbations, we may arrive at the two-point correlation function $\Delta_\mathcal{R}\equiv \left\vert v_{\mathcal{R},k} \right\vert^2$, which is
given by
\begin{equation}
   \label{powspecscalsdim}
\Delta_\mathcal{R} \   =\ \frac{H^4}{Z_\mathcal{R}   \dot \varphi^2}\; ,
\end{equation}
where from now on, we use unbarred quantities to denote the background.

Observable cosmological quantities which can be measured on the CMB are linked to the primordial perturbations through transfer functions which induce a multiplicative multipole contribution to the power spectrum of scalar perturbations \cite{Turner:1993vb},
\begin{equation}
    \label{powspecscal}
P_\mathcal{R}\ \equiv\ \frac{k^3}{4\pi^2}\,\Delta_\mathcal{R}\; .
\end{equation}
In a similar way, we may write the power spectrum $P_T$ for tensor perturbations calculated through the correlation function $\Delta_T \equiv \left\vert v_{T,k} \right\vert^2 = H^2/Z_T$, i.e.
\begin{equation}
\label{powspectens}
P_T\ =\ \frac{2 k^3 }{ \pi^2}\, \Delta_T\; .
\end{equation}
In the standard Einstein gravity, \eqref{powspecscal} and \eqref{powspectens} reduce to their usual expressions. The power spectra are further related to the primordial density perturbation via a multiplicative factor due to quadrupole anisotropies. However, the scale dependence of the spectrum is independent of this factor. The scale dependence, which is termed the {\it spectral index} or {\it scalar tilt}, is defined as 
\begin{equation}
   \label{nR}
n_\mathcal{R}\: -\: 1\ \equiv\ \left.\frac{d \ln  \Delta_\mathcal{R} }{d\ln k}\right\vert_{k = aH}\;,
\end{equation}
where $k = aH$ is the horizon crossing condition, since the horizon crossing time is when the perturbations left their observable imprint on the scalar tilt. There is an analogous relation for the {\it tensor tilt},
\begin{equation}
   \label{nT}
n_T \ \equiv\ \left.\frac{d \ln  \Delta_T }{d\ln k}\right\vert_{k = aH}\; .
\end{equation}
Another useful observable is the \emph{tensor-to-scalar ratio} $r$, which is defined as
\begin{equation}
   \label{str}
r\ \equiv\    \frac{  P_T }{  P_\mathcal{R} }  \ .
\end{equation}
Finally, we may define the \emph{running} of the spectral indices, which encodes their scale dependence, as follows:
\begin{eqnarray}
\label{runRdef}
\alpha_\mathcal{R}  &\equiv& \left.\frac{d    n_\mathcal{R} }{d\ln k}\right\vert_{k = aH}\; ,\\
\label{runTdef}
\alpha_T  &\equiv& \left.\frac{d    n_T }{d\ln k}\right\vert_{k = aH}\; .
\end{eqnarray}
Observe that the power spectra and all observable quantities derived from them depend solely on the background, even though these quantities are of pure quantum-mechanical origin.

\section{Slow-Roll Inflation}\label{sec:srinf}

In this section we present the \emph{slow-roll approximation} formalism which is often employed to approximate the equations of motion governing the inflationary dynamics in scalar-curvature theories, as well as calculate all cosmological observables from the scalar and tensor power spectra~$P_{\mathcal{R}}$ and~$P_T$~[cf.~\eqref{nR} and \eqref{str}]. To this end, we first define the Hubble slow-roll parameters which we use to express the predictions for the cosmological observables of inflation, such as $n_{\cal R}$, $n_T$ and~$r$. We then discuss the inflationary attractor solution, with aid of which the inflationary observables can be expressed in a concise manner in terms of new {\em potential slow-roll parameters} which only depend on the model functions $f(\varphi )$, $k (\varphi )$, and $V(\varphi )$ and their derivatives with respect to~$\varphi$.

\subsection{Hubble Slow-Roll Inflation}

The basic working hypothesis in the slow-roll approximation formalism is that, during inflation, the following double inequality holds:
\begin{equation}
    \label{srap}
\ddot \varphi\ \ll\ H \dot \varphi\ \ll\ H^2 \varphi\; .
\end{equation}
The above hierarchy of energy scales was first considered to describe  minimal inflation~\cite{liddle94}. Nevertheless, this hierarchy of scales can be extended to non-minimal inflation in general scalar-curvature theories by noting that, for any well-behaved function $g(\varphi)$ of the inflaton $\varphi$, one may require~\cite{Torres:1996fr} that
\begin{align}
   \label{srg}
\ddot g\  \ll\ H \dot g\  \ll\ H^2 g\; .
\end{align}
This motivates us to  define the following \emph{Hubble slow-roll parameters}~\cite{noh01}\footnote{Note that our notation for these parameters can be linked to the notation in \cite{noh01} by 
\begin{equation*}
\epsilon_H\ =\ -\epsilon_1\;,\qquad 
\delta_H\ =\ -\epsilon_2\; ,\qquad 
\kappa_H\ =\ \epsilon_3\; ,\qquad 
\sigma_H\ =\ \epsilon_4\; .
\end{equation*}
}:
\begin{align}
   \label{epsdef}
\epsilon_H\ &\equiv\ -\frac{\dot H}{H^2}\ ,
& \delta_H\ &\equiv\ -\,\frac{\ddot \varphi}{H \dot \varphi}\ ,& \\
   \label{kappadef}
\kappa_H\ &\equiv\ \frac{1}{2} \frac{\dot f}{H f}\ =\ \frac{1}{2}\frac{f_{,\varphi} \dot \varphi}{Hf}\ ,
& \sigma_H\ &\equiv\ \frac{1}{2} \frac{\dot E}{H E}\ =\ \frac{1}{2}\frac{E_{,\varphi} \dot \varphi}{EH}\ ,&
\end{align}
with
\begin{equation}
   \label{Edef}
E \ \equiv\  kf\: +\: \frac{3}{2}f_{,\varphi} ^2\; .
\end{equation}
Note that the Hubble slow-roll parameters provide a measure of the deviation of the universe from an exact de Sitter space. In particular, the slow-roll parameters $\kappa_H$ and $\sigma_H$ defined in~\eqref{kappadef} are necessary to fully describe the dynamics in non-minimal inflation. In the Einstein frame, it is also possible to establish the relation $\eta_H \equiv \dot \epsilon_H/(H\epsilon_H)
= 2\epsilon_H - 2\delta_H$. However,  for the scalar-curvature theories under study, $\delta_H$ proves to be more convenient for computing cosmological observables of inflation~\cite{Stewart:1993bc}, rather than the more frequently used parameter~$\eta_H$. For later convenience, we also define the quantity
\begin{equation}
  \label{baretaH}
\overline{\eta}_H\ \equiv\ \epsilon_H\: +\: \delta_H\;,
\end{equation}
which differs from $\eta_H$.

We are now in a position to calculate the cosmological observables of inflation in the slow-roll approximation in terms of the Hubble slow-roll parameters defined in~\eqref{epsdef} and~\eqref{kappadef}. With the help of these parameters, we may write down $Z_\mathcal{R}$ given in~\eqref{smeqdef} as
\begin{align}\label{ZRdef}
Z_\mathcal{R}\ =\  \frac{k + \frac{3 \dot f^2}{2f\dot \varphi^2} }{\left( 1+ \frac{\dot f}{2Hf}\right)^2 }\ =\ \frac{E/f}{(1+\kappa_H)^2}\ .
\end{align}
Observe that $Z_{\cal R}$ is a key parameter in inflationary dynamics, as it enters the definition of the scalar power spectrum~$P_{\cal R}$ in~\eqref{powspecscal} through $\Delta_\mathcal{R}$ in~\eqref{powspecscalsdim}. 

Since the cosmological perturbations freeze outside the horizon, their power spectrum $P_{\cal R}$ when they first exit the horizon will match the one observed at the present epoch, assuming that we only observe scales that are just re-entering the horizon without having time to evolve further. The condition for this {\em second} horizon crossing due to the re-entry of the perturbations is given by~$aH = k$. The latter can be rewritten as
\begin{align}
   \label{eq:lnk}
\ln k\ =\ \ln a\: +\: \ln H\; .
\end{align}
Employing the definition of \emph{e-folds}: $dN = H d\tau = d\ln a$ and the relation~\eqref{eq:lnk}, we easily find that
\begin{align}
   \label{dlnkdN1}
\frac{d\ln k}{dN}\ =\ 1\: +\: \frac{d\ln H}{dN}\ =\ 1\: -\: \epsilon_H\; .
\end{align}
We may now calculate the spectral index $n_{\cal R}$ in~\eqref{nR}, by using the chain rule,
by means of~\eqref{dlnkdN1}, along with the expression for $Z_\mathcal{R}$ in~\eqref{ZRdef}. Keeping only the leading order in a series expansion of the slow-roll parameters, we arrive at~\cite{Hwang:1996xh, noh01}
\begin{align}
    \label{unified}
n_\mathcal{R}\ =\ 1\: -\: 4\epsilon_H\: +\: 2\delta_H\: +\: 2 \kappa_H\:  -\: 2\sigma_H\; .
\end{align}
In deriving~\eqref{unified}, we have assumed that the Hubble slow-roll parameters are slowly varying, such that we may discard their time derivatives, as dictated by the generalized slow-roll approximation given in~\eqref{srg}. Proceeding as above for the cosmological observables $n_T$ and $r$ defined in~\eqref{nT} and~\eqref{str}, respectively, we obtain in the slow-roll approximation,
\begin{align}
\label{nTH}
n_T\ =\ -2\epsilon_H\: -\: 2 \kappa_H\; ,\\
\label{tenstoscalrat}
r \ =\  16  \epsilon_H\: +\:  16 \kappa_H  \; .
\end{align}
Analogous leading-order expressions can be derived for $\alpha_\mathcal{R}$ and $\alpha_T$, by using~\eqref{unified} and~\eqref{nTH} in their definitions \eqref{runRdef} and \eqref{runTdef}.

In spite of having expressed the inflationary observables in a compact form in terms of the Hubble slow-roll parameters, their accurate evaluation at horizon crossing remains still a challenge. In particular, the inflationary observables often depend crucially on the number of e-folds, given by~$N(t,t_\text{end}) = \int_t^{t_{\rm end}} H\,dt$ (with $N_L  =1$), where the cosmic time $t_{\rm end}$ characterizing the end of inflation may be determined by the condition
\begin{align}
\text{max}(\epsilon_H, |\overline{\eta}_H |)\ =\  1\; ,
\end{align}
where $\overline{\eta}_H$ is given in~\eqref{baretaH}.  However, this condition only applies in the Einstein frame, and as we will show in~Section~\ref{sec:reparam}, it is frame-{\em dependent}, which necessitates the introduction of an appropriate frame-covariant extension for it to be applicable in general Jordan frames.

\subsection{The Inflationary Attractor Trajectory}

In the Einstein frame, the equations of motions can be drastically simplified, if certain conditions are met, which assure that the slow-roll parameters are sufficiently small at early times, so as to successfully generate inflationary dynamics.  These conditions select a class of solutions known as the \emph{inflationary attractor trajectory}, to which all other inflationary trajectories converge rapidly independent of their initial position in phase space, such that the use of the so-called slow-roll approximation is justified~\cite{Faraoni:2000nt}.  As we will see in the next subsection, we may use the inflationary attractor solution to write the Hubble slow-roll parameters in terms of the potential~$U(\varphi )$ and its derivatives with respect to the inflaton field~$\varphi$. Thus, one may define a new set of fully equivalent parameters called the \emph{potential slow-roll parameters}. Our aim is to generalize this procedure to scalar-curvature theories in the Jordan frame.

The first step in doing so is to derive the approximate equations of motion that govern the inflationary attractor trajectory. Therefore, we start with  the generalized equations of motion for the background metric and inflaton fields, and express them in terms of the Hubble slow-roll parameters $\epsilon_H$, $\delta_H$, $\kappa_H$ and $\sigma_H$. With the help of \eqref{epsdef} and \eqref{kappadef}, and upon substitution of~\eqref{eomj3} into~\eqref{eomj2}, the Friedmann equation may be rewritten as
\begin{equation}
   \label{Hequat}
 H^2\ =\  \frac{fU}{3}     \left(1 \:  -\: \frac{\epsilon_H}{3}\:   -\:  \frac{\theta_H}{3} \right)^{-1} \,.
\end{equation}
For convenience, we have defined a new slow-roll parameter $\theta_H$,
\begin{equation}
   \label{thetadef}
\theta_H\ \equiv\ \frac{1}{2}\,\frac{\ddot f}{  H^2   f}\ .
\end{equation}
As we will now show, $\theta_H$ is of higher order than the rest of the slow-roll parameters.
Using \eqref{kappadef} in \eqref{thetadef}, it follows that
\begin{equation}
   \label{thetarel}
\theta_H\ =\ \frac{1}{2}\left(\frac{f_{,\varphi\varphi}  \dot \varphi^2}{H^2 f}\: -\: \delta_H \kappa_H \right)\; .
\end{equation}
In order to eliminate the term $\propto f_{,\varphi\varphi}$ in~\eqref{thetarel}, we
use the fact that  $\dot \kappa_H/H$ can also be written as
\begin{equation}
   \label{kappadot} 
\frac{\dot \kappa_H}{H}\ =\
 \frac{f_{,\varphi\varphi} \dot\varphi^2   }{H^2f}\: -\: \delta_H\kappa_H 
\: +\: \epsilon_H\kappa_H  
\: -\: \kappa_H^2\; .
\end{equation}
Using \eqref{kappadot} in \eqref{thetarel}, we finally arrive at
\begin{equation} 
    \label{thetaeq}
\theta_H\ =\ \frac{1}{2}\left( \frac{\dot \kappa_H}{H \kappa_H}\: -\: \epsilon_H\: +\: \kappa_H \right)\kappa_H\;.
\end{equation}
The latter shows that, to the leading {\em linear} order in the slow-roll approximation, $\theta_H$ vanishes.
This result is also consistent with the one that one would have naively obtained by considering the double inequality in~\eqref{srg}.

Similarly, after dividing~\eqref{eomj1} by $3H\dot \varphi E/f$, where $E$ is defined in~\eqref{Edef}, the inflaton equation of motion may be recast into the form,
\begin{equation}
   \label{infl}
 1\: -\:  \frac{1}{ 3 }  \delta_H\: 
 +\: \frac{2}{3 } \sigma_H\:
 =\ -\, \frac{f^2 U_{,\varphi}}{ 3H{\dot \varphi} E/f  }\; .
\end{equation}
Finally, the acceleration equation \eqref{eomj3}  can be written down as
\begin{equation}
   \label{contsr}
\epsilon_H\:  +\:  \kappa _H\: -\: \theta_H\ =\   \frac{k \dot \varphi^2  }{2H^2 f}\; .
\end{equation}

The inflationary attractor solution is obtained by considering only the leading terms in the 
Friedmann and inflaton equations \eqref{Hequat} and \eqref{infl}. Hence, ignoring all terms 
depending on the Hubble slow-roll parameters, we find that \eqref{Hequat} and \eqref{infl}
simplify to
\begin{equation}
    \label{IAT}
 H^2\ \approx\  \frac{ f\,U}{3}\;,\qquad
H {\dot \varphi}\ \approx\ -\,\frac{f^3\, U_{,\varphi}}{3\,E}\; ,
\end{equation}
which determine the inflationary attractor trajectory in the leading slow-roll approximation. By dividing separately the LHSs and RHSs of the two equations of motions in~\eqref{IAT}, we obtain the useful relation
\begin{equation}
    \label{dotphiH}
\frac{ H }{{\dot \varphi} }\ \approx\ -\,\frac{E\, U }{f^2\, U_{,\varphi} }\; .
\end{equation}
With the aid of~\eqref{IAT} and~\eqref{dotphiH}, in the next subsection we can define a new set of potential slow-roll parameters, which will be used to express all relevant cosmological observables of inflation in a concise manner.

\subsection{Potential Slow-Roll Inflation}
\label{subsec:potsrinf}

Having derived the equations of motion that determine the inflationary attractor trajectory in~\eqref{IAT}, we are now in a position to express the Hubble slow-roll parameters $\epsilon_H$, $\delta_H$, $\kappa_H$ and $\sigma_H$, given in~\eqref{epsdef}, in terms of the model functions $f(\varphi )$, $k (\varphi )$ (or $E(\varphi )$) and $U(\varphi )$ and their derivatives with respect to~$\varphi$, without making any explicit reference to the Einstein frame. The new parameters so derived will be called the {\it potential slow-roll parameters} to distinguish them from the Hubble slow-roll parameters in~\eqref{epsdef} and they will be valid in any Jordan frame.

We start our derivation by noticing that time derivatives acting on $\varphi$, e.g.~$\dot\varphi$, can be eliminated by virtue of~\eqref{dotphiH} and that time derivatives acting on the Hubble parameter $H$, e.g.~$\dot H$, may also be replaced with $\varphi$-derivatives acting on $f$ and $U$, after differentiating both sides of the first equation in~\eqref{IAT} with respect to the rescaled cosmic time~$\tau$. In this way, we may derive from the Hubble slow-roll parameters a new set of parameters, such that $\epsilon_U \approx \epsilon_H$, $\delta_U \approx \delta_H$, $\kappa_U \approx \kappa_H$ and $\sigma_U\approx \sigma_H$, which do {\em not} depend on~$H$ and $\dot\varphi$.

Following carefully the procedure mentioned above, the potential slow-roll parameters generalized in the Jordan frame are found to be
\begin{equation}
\begin{aligned}
 \epsilon_U\
&\equiv\ \frac{1}{2}\, \frac{f  U_{,\varphi} (f U)_{,\varphi}}{EU^2}\ ,
&
\qquad
\delta_U\  
&\equiv\ 
 \frac{1}{2}\, \frac{f  U_{,\varphi} (f U)_{,\varphi}}{EU^2}\:
 +\: \left(\frac{f^2 U_{,\varphi}}{ E U} \right)_{,\varphi}\ ,
\\
\label{kappaU}
\kappa_U\ &\equiv\  -\,\frac{f_{,\varphi}  }{2  }\, \frac{f  U_{,\varphi} }{EU  }\ ,
&
\qquad
\sigma_U\ &\equiv\ - \,\frac{1}{2}\,\frac{E_{,\varphi} }{E^2   }\,  \frac{f^2 U_{,\varphi} }{U }\ .
\end{aligned}
\end{equation}
Because of the equivalence between the Hubble and potential slow-roll parameters in the slow-roll approximation, writing down the analytical formulae for the tensor-to-scalar ratio~$r$, and the spectral indices $n_\mathcal{R}$ and $n_T$ in terms of the latter parameters becomes a simple task. In fact, what we only need to do is to replace the Hubble slow-roll parameters in the expressions~\eqref{tenstoscalrat}, \eqref{unified} and \eqref{nTH}, with their potential counterparts: \begin{eqnarray}
   \label{tenstoscalratpot}
r &=&  16 \epsilon_U\: +\:  16 \kappa_U  \; ,\\
   \label{unifiedpot}
n_\mathcal{R} &=& 1\: -\: 4\epsilon_U\: +\: 2\delta_U\: +\: 2 \kappa_U\:  -\: 2\sigma_U\;,\\
   \label{ntt}
n_T &=& -\,2\epsilon_U\: -\: 2 \kappa_U\; .
\end{eqnarray}
Note that the results for $r$ and $n_T$ confirm the so-called ``consistency relation'' of minimal inflation,
\begin{equation}
   \label{conrel}
r \ =\ -\,8n_T\; ,
\end{equation}
which remains also valid in the context of general scalar-curvature theories. 

We may similarly proceed to derive analytic expressions for the runnings $\alpha_\mathcal{R}$ and $\alpha_T$ of the spectral indices $n_\mathcal{R}$ and $n_T$ in terms of the potential slow-roll parameters.
With this aim, we first note the useful chain-rule relation:
\begin{equation}
  \label{dphidN}
  \frac{d\varphi}{d\ln k}\ =\ \frac{d\varphi}{d N}\, \frac{dN}{d\ln k}\ =\ \frac{\dot\varphi}{H}\, 
  (1 - \epsilon_H)^{-1} \ \approx\ -\, \frac{f^2\,U_{,\varphi}}{E\,U}\ .
\end{equation}
In arriving at the last expression in~\eqref{dphidN}, we first used~\eqref{dlnkdN1} and then~\eqref{dotphiH}, and approximated $(1 - \epsilon_H)^{-1} \approx 1$, in the leading slow-roll approximation.  On the basis of the definitions for the spectral runnings~$\alpha_\mathcal{R}$ and~$\alpha_T$ in~\eqref{runRdef} and~\eqref{runTdef}, and after employing the chain-rule relation~\eqref{dphidN}, we find
\begin{eqnarray}
   \label{runscal}
\alpha_\mathcal{R} &=& -\,\frac{f^2\, U_{,\varphi}}{E\,U}\, {n_\mathcal{R}}_{ ,\varphi} 
\ =\ \frac{f^2 U_{,\varphi}}{E\,U}\: \Big( 4\epsilon_U\: -\: 2\delta_U\: -\: 2 \kappa_U\: 
+\: 2\sigma_U  \Big)_{,\varphi}\;,\\
   \label{runtens}
\alpha_T &=&  -\,\frac{f^2\, U_{,\varphi}}{E\,U}\, {n_T}_{ ,\varphi} 
\ =\ \frac{f^2\, U_{,\varphi}}{E\,U}\: \Big( 2\epsilon_U\: +\: 2\kappa_U\Big)_{,\varphi}\;.
\end{eqnarray}

In Section~\ref{sec:specmod}, we will use the analytical expressions stated in~\eqref{tenstoscalratpot}, \eqref{unifiedpot}, \eqref{runscal} and \eqref{runtens} to obtain predictions for all relevant cosmological observables of inflation in specific models. In this context, we should remark here that all the inflationary observables of interest must be evaluated at inflaton field values $\varphi$, which typically correspond to the time when the observed cosmological scales have left the horizon, i.e.~about $N = 60$  e-folds before the end of inflation.

Another advantage of our formalism is that the four potential slow-roll parameters $\epsilon_U$, $\delta_U$, $\sigma_U$ and $\kappa_U$ and their $\varphi$-derivatives suffice to calculate all the observables to leading order in the slow-roll approximation. Indeed, if we were to calculate higher runnings of the spectral indices, we would not need to introduce new slow-roll parameters as is usually done in the Einstein frame, but only higher derivatives of $\epsilon_U$, $\delta_U$, $\sigma_U$ and $\kappa_U$ with respect to $\varphi$. Nonetheless, we may confirm that our expressions for the cosmological observables reduce to the well-known ones quoted in the literature for the minimally coupled inflation models~\cite{Bassett:2005xm}: 
\begin{align}
   \label{minimalobs}
r\ &=\ 16 \epsilon_V\; ,
& n_\mathcal{R}\  &=\ 1\: -\: 6\epsilon_V\: +\: 2\eta_V\;,
&
n_T\ &=\  -2\epsilon_V\;,
&
\nonumber \\
& & \alpha_\mathcal{R}\ &=\  16\epsilon_V \eta_V\: -\:  24 \epsilon_V^2\: -\: 2\xi_V^2\;,
&
\alpha_T\ &=\ -8\epsilon_V^2\: +\: 4 \epsilon_V \eta_V\;,
\end{align}
which are expressed in terms of the Einstein-frame parameters $\epsilon_V, \eta_V$, and $\xi_V$, given by
\begin{align} 
\epsilon_V\ &\equiv\ \frac{M_P^2}{2} \frac{V^2_{,\varphi}}{V^2}\ ,
\\
\eta_V\ &\equiv\  \frac{M_P^2 V_{,\varphi\varphi}}{V}\ =\ \epsilon_V\: +\: \delta_V,
\\
\xi_V^2\ &\equiv\  \frac{M_P^4  V_{,\varphi} V_{,\varphi\varphi\varphi}}{V^2}\ .
\end{align}

For completeness, we also derive simplified expressions for the power spectra of the curvature and tensor perturbations $P_\mathcal{R}$ and $P_T$, given by \eqref{powspecscal} and \eqref{powspectens}, in the slow-roll approximation. Employing \eqref{ZRdef}, \eqref{IAT}, and \eqref{dotphiH},  the power spectra $P_\mathcal{R}$ and  $P_T$ take on the simple form
\begin{eqnarray}
P_\mathcal{R}\ &\approx&  \frac{k^3}{12\pi^2 } \frac{E\,U^3}{f^2\, U_{,\varphi}^2}\ =\ \frac{k^3}{24\pi^2 }\, \frac{U}{\epsilon_U + \kappa_U}\ , \\
P_T\ &\approx& \frac{2 k^3 }{3\pi^2}\, U\; .
\end{eqnarray}
Finally, thanks to~\eqref{dotphiH}, the number of e-folds may be evaluated to leading order in 
the slow-roll approximation as
\begin{equation} 
   \label{Nphi}
N (\varphi)\ =\ -\int_{\varphi}^{\varphi_\text{end}}  d\varphi' \, \frac{E (\varphi') }{f (\varphi')^2 }\,  \frac{ U (\varphi')  }{U (\varphi')_{ ,\varphi'}}\ ,
\end{equation}
where $\varphi_\text{end}$ is the inflaton value at the end of inflation, which is usually determined by the condition:
\begin{equation} 
   \label{maxEF}
\max(\epsilon_U, |\eta_U|)\ =\ 1\; ,
\end{equation}
with $\eta_U \equiv \epsilon_U +\delta_U$. 

An obstacle in our approach to derive a frame-covariant formulation of inflation is the fact that the number of e-folds~$N$, and especially $\varphi_\text{end}$, are {\em not} frame-invariant quantities. This means that the end-of-inflation condition~\eqref{maxEF} requires a non-trivial extension in order to hold true in an arbitrary Jordan frame.  In order to be able to find this missing piece of information, we study in the next section the transformation properties of the potential slow-roll parameters under frame transformations.

\section{Frame Covariance}\label{sec:reparam}

In this section, we will use the results that we derived in Section~\ref{sec:srinf}, in the context of scalar-curvature theories, in order to evaluate the cosmological observables in different frames. Even though the action~$S$ for these theories is not invariant under conformal rescalings of the metric~$g_{\mu\nu}$ and field reparameterizations of the inflaton field~$\varphi$, its functional dependence on the transformed metric~$\tilde{g}_{\mu\nu}$ and inflaton field~$\tilde{\varphi}$, and the transformed model functions~$\tilde{f}$, $\tilde{k}$ and $\widetilde{V}$, does not change [cf.~\eqref{actiontransprop}]\footnote{Our frame-covariant approach to scalar-curvature theories is general, as it describes the transformation properties of 
kinematic parameters from one Jordan frame to another {\em arbitrary} Jordan frame. As such, it includes the special class of Jordan-to-Einstein frame transformations discussed in~\cite{Postma:2014vaa}, {\em without} making any \emph{a priori} assumptions about the frame invariance (or lack thereof) of scalar-curvature theories as in~\cite{Dicke:1961gz}.}.  Given the transformation properties of these quantities, we may determine how the potential slow-roll parameters transform. Taking the latter into account, we will show that the physical cosmological observables remain invariant under frame transformations in the leading order of the slow-roll approximation.

\subsection{Conformal Transformations}

Let us first examine how the generalized potential slow-roll parameters~$\epsilon_U$, $\delta_U$, $\kappa_U$ and~$\sigma_U$, as defined in~\eqref{kappaU}, would change by considering only a conformal rescaling of the metric~$g_{\mu\nu }$, according to~\eqref{weyl}.  For this purpose, we take into account the relations of the transformed model functions $\tilde f$, $\tilde k$ and $\widetilde V$  in terms of the original ones $f$, $k$ and $V$, by setting $K=1$ in \eqref{Imodelparamdef}, which amounts to $\tilde{\varphi} = \varphi$. We then find that the quantities $U$ and $E$, given in~\eqref{Udef} and~\eqref{Edef}, transform correspondingly as 
\begin{equation}
   \label{Etranst}
\widetilde U (\varphi)\  =\ U(\varphi)\;, 
\qquad
\widetilde E (  \varphi)\ =\ \frac{E (\varphi )}{\Omega^4}\ .
\end{equation}
We may now use the original definitions of the slow-roll parameters in~\eqref{kappaU}
to compute the transformed ones, by replacing $U(\varphi) \to \widetilde{U}(\tilde\varphi)$ and 
$f(\varphi ) \to \tilde{f}(\tilde\varphi )$, with $\tilde{\varphi} = \varphi$. In this way, we have, for example,
\begin{equation}
   \label{epsilonUtild}
\tilde \epsilon_U (  \varphi)\ =\ \frac{1}{2} \frac{\tilde f  \widetilde U_{,\tilde \varphi} (\tilde f \widetilde U)_{,\tilde \varphi}}{\widetilde E\widetilde U^2}\ ,
\end{equation}
and similarly for $\tilde \delta_U$, $\tilde \kappa_U$ and $\tilde \sigma_U$. If we expand the new slow-roll parameters by means of~\eqref{Imodelparamdef}, we find that they transform as
\begin{align}
    \label{upatrans}
\tilde \epsilon_U  (\varphi)\ &=\ \epsilon_U (\varphi)\: -\: \Delta_\Omega (\varphi)\;,
&
\tilde \delta_U (\varphi)\  &=\ \delta_U (\varphi)\: -\: \Delta_\Omega (\varphi)\;,
\nonumber\\
\tilde \kappa_U (\varphi)\ &=\ \kappa_U (\varphi)\: +\:  \Delta_\Omega (\varphi)\;,
&
\tilde \sigma_U (\varphi)\  &=\ \sigma_U (\varphi)\:  +\: 2\Delta_\Omega (\varphi)\; ,
\end{align}
where
\begin{equation}
   \label{deltaOmega}
\Delta_\Omega\ \equiv\  \frac{ f^2 U_{,\varphi}  }{E U}\,\frac{\Omega _{,\varphi}}{\Omega}\ .
\end{equation}
Evidently, depending on the actual value of~$\Delta_\Omega$, the slow-roll parameters may not be small and can have either sign after a conformal transformation.

\subsection{Inflaton Reparameterizations}

Let us now discuss a second class of general frame transformations, under which only the inflaton field $\varphi$ gets reparameterized as $\varphi \to \tilde{\varphi} = \tilde{\varphi}(\varphi )$. Such a field reparameterization may be determined through the differential equation $\left( d \tilde \varphi/d\varphi \right)^2 = K (\varphi)$~[cf.~\eqref{Kphi}], where $K (\varphi)$ is an arbitrary function of~$\varphi$.  Inflaton reparameterizations, with $K (\varphi ) = k(\varphi )$, are usually performed in the literature to make the inflaton kinetic term canonical, but here we will not impose this restriction.

Applying~\eqref{Imodelparamdef} to the transformed model functions $\tilde{f}$, $\tilde{k}$ and $\widetilde{V}$ for $\Omega = 1$  yields
\begin{equation}
\tilde   f(\tilde \varphi)\ =\ f(\varphi(\tilde \varphi))\; ,\qquad
\tilde   k(\tilde \varphi)\ =\ \frac{k(\varphi(\tilde \varphi))}{K(\varphi(\tilde \varphi))}\ ,\qquad
\tilde   V(\tilde \varphi) \ =\ V(\varphi(\tilde \varphi))\; ,
\end{equation}
implying that 
\begin{equation}
    \label{Etrans}
\widetilde E(\tilde \varphi) \
=\ \frac{E(\varphi (\tilde \varphi))}{K(\varphi (\tilde \varphi))}\ .
\end{equation}
Here, we have assumed that the function $\tilde{\varphi} = \tilde{\varphi}(\varphi)$
can be inverted to $\varphi = \varphi(\tilde \varphi)$, at least piecewise.

As was done above for the case of conformal transformations only, we rely on the analytical expressions given in~\eqref{kappaU} to calculate the transformed slow-roll parameters $\tilde \epsilon_U$, $\tilde \delta_U$, $\tilde \kappa_U$ and $\tilde \sigma_U$, as functions of $\tilde\varphi$. We then use the chain rule, 
\begin{equation} 
\frac{d}{d\tilde\varphi}\ =\ \frac{1}{\sqrt{K(\varphi)}} \frac{d}{d\varphi} 
\end{equation} 
to re-express them in terms of the original slow-roll parameters $\epsilon_U$, $\delta_U$, $\kappa_U$ and $\sigma_U$, which depend on the original inflaton field $\varphi$.
Thus, under an inflaton reparameterization, the slow-roll parameters transform as 
\begin{align}
    \label{epsilonUhat} 
\tilde \epsilon_U (\tilde \varphi )\ &=\ \epsilon_U (\varphi)\; , & 
\tilde \delta_U (\tilde \varphi)\ &=\ \delta_U ( \varphi)\: +\: \Delta_K ( \varphi)\; ,\nonumber\\ \tilde \kappa_U (\tilde \varphi)\ &=\ \kappa_U ( \varphi)\; , & \tilde \sigma_U (\tilde \varphi)\ &=\ \sigma_U ( \varphi)\: +\: \Delta_K ( \varphi)\; , 
\end{align} 
where we have defined 
\begin{equation}
    \label{deltaK}
\Delta_K (\varphi)\ \equiv\ \frac{1}{2} \frac{ K_{,\varphi}}{K  } \frac{f  ^2 U  _{,\varphi}}{ E    U  }\ .
\end{equation}
Notice that under a reparameterization of the inflaton field~$\varphi$, only the slow-roll parameters~$\delta_U$ and $\sigma_U$ get transformed.

\subsection{Invariance Under Frame Transformations}\label{sec:invframe}

Given the transformation properties of the potential slow-roll parameters stated in~\eqref{upatrans}  and~\eqref{epsilonUhat}, it is straightforward to show that the cosmological observables of inflation, such as the tensor-to-scalar ratio $r$, the scalar and tensor spectral indices $n_{\cal R}$ and $n_T$, and their runnings $\alpha_{\cal R}$ and $\alpha_T$, do {\em not} depend on the choice of frame in the leading order of the slow-roll approximation.  Employing the analytical expressions~\eqref{tenstoscalratpot}, \eqref{unifiedpot}, \eqref{ntt}, \eqref{runscal} and \eqref{runtens} for the aforementioned 
cosmological observables in terms of slow-roll parameters, we find that
\begin{align}
    \label{obinvar}
\widetilde r  (\tilde \varphi)\ &=\ r(\varphi)\;,
&
\widetilde n_T  (\tilde \varphi)\ &=\   n_T (\varphi)\;,
&
\widetilde n_\mathcal{R} (\tilde \varphi)\  &=\ n_\mathcal{R} (\varphi)\;,
\nonumber\\
& & \widetilde \alpha_\mathcal{R}  (\tilde \varphi)\ &=\ \alpha_\mathcal{R}(\varphi)\;,
& 
\widetilde \alpha_T  (\tilde \varphi)\ &=\   \alpha_T(\varphi)\; .
\end{align}
It should be stressed here that the inflationary observables $r$, $n_{\cal R}$, $n_T$, $\alpha_{\cal R}$ and $\alpha_T$ are invariant under the {\em separate} action of conformal rescalings of the metric~$g_{\mu\nu}$ and field reparameterizations of the field~$\varphi$. 

The frame invariance of the cosmological observables shown above holds, as long as their
$\varphi$-dependence through $\varphi =  \varphi (\tilde{\varphi})$ is taken into account.
However, this frame invariance is spoiled, once the same observables are naively expressed
in terms of the number of e-folds $N$. In fact, under conformal rescalings of the metric, the number of e-folds $N$ does transform and is {\em not} frame-invariant. To see this explicitly, 
we first note that the exact determination of $N$ is given by
\begin{align} 
N \ &=\   \int^{t_\text{end}}_{t}  N_L H\,dt' \ =\ \int^{a_\text{end}}_{a} \, \frac{da'}{a'}\ =\ \ln\left(\frac{a_\text{end}}{a}\right),
\end{align}
which transforms to
\begin{align} 
   \label{Ntilde}
\widetilde N\ &=\ \int^{\tilde a_\text{end}}_{\tilde a} \, \frac{d\tilde a'}{\tilde a'}\
=\
\int^{a_\text{end}}_{a} \, \frac{  da'   }{  a'}\
+\ \int^{\Omega_\text{end}}_{\Omega} \, \frac{    \, d\Omega }{\Omega  }\
=\
N\: +\: \ln \left(\frac{\Omega_\text{end}}{\Omega}\right)\; .
\end{align}
Thus, $N$ is not frame invariant, as it receives an extra contribution given by $\ln (\Omega_\text{end}/\Omega)$.  It is interesting to compare this last result with the corresponding one that would have been obtained by virtue of~\eqref{Nphi} which was derived in the slow-roll approximation.  Making use of the transformation properties of $E(\varphi)$ and $f(\varphi)$ reported above, we find that the integrand in~\eqref{Nphi}
remains unaltered under frame transformations. However, the field value of $\varphi$ at the end of inflation, $\varphi_{\rm end}$, is usually determined by the condition $\max(\epsilon_U, |\eta_U|) = 1$
[cf.~\eqref{maxEF}], which is only applicable in the Einstein frame. Since the end-of-inflation condition~\eqref{maxEF} is frame-{\em dependent}, we need to deduce its frame-invariant generalization that should hold to {\em any}  Jordan frame. This is given by
\begin{align}
   \label{maxJF}
\max(\epsilon_U+\kappa_U, |\epsilon_U + \delta_U +4\kappa_U - \sigma_U | )\ =\ 1\; .
\end{align}
As we will show below, this generalization is unique and reduces to the Einstein case when $\kappa_U = \sigma_U = 0$. Hence, by means of~\eqref{maxJF}, we have $\tilde \varphi_\text{end} =   \tilde{\varphi}(\varphi_\text{end})$, and so $N(\varphi) = \widetilde{N} (\tilde{\varphi})$ in the slow-roll approximation. This exercise tells us that the
unwanted term  $\ln (\Omega_\text{end}/\Omega)$ on the RHS of~\eqref{Ntilde} would
only become significant beyond the leading order of the slow-roll approximation.
Consequently, we have shown that all relevant cosmological observables are frame-invariant 
when expressed in terms of the number of e-folds $N$ in the slow-roll approximation:
\begin{align}
    \label{obinvar2}
\widetilde r  (\widetilde N)\ &=\ r(N)\; ,
&
\widetilde n_T  (\widetilde N)\ &=\   n_T (N)\; ,
&
\widetilde n_\mathcal{R} (\widetilde N)\ &=\ n_\mathcal{R} (N)\; ,
\nonumber\\
& &\widetilde \alpha_\mathcal{R}  (\widetilde N)\ &=\ \alpha_\mathcal{R}(N)\; ,
&
\widetilde \alpha_T  (\widetilde N)\ &=\   \alpha_T(N)\; .
\end{align}

In order to explicitly demonstrate the uniqueness of the end-of-inflation condition~\eqref{maxJF}, we will prove that demanding $\sigma_U = \kappa_U = 0$ uniquely singles out the Einstein frame. With the help of~\eqref{upatrans} and \eqref{epsilonUhat}, 
we readily see that $\sigma_U$ and $\kappa_U$ transform as 
\begin{equation} 
\tilde \kappa_U\ =\ \kappa_U\: +\: \Delta_\Omega\;,\qquad
\tilde \sigma_U\ =\  \sigma_U\: +\: 2\Delta_\Omega\: +\: \Delta_K\; .  
\end{equation} 
Requiring the vanishing of $\tilde \sigma_U$ and $\tilde \kappa_U$  implies
\begin{equation}
  \Delta_\Omega\ =\ -\,\kappa_U\;,\qquad   \Delta_K\ =\ -\,\sigma_U\: +\: 2\kappa_U\; .
\end{equation}
Using the definition of $\Delta_\Omega$ and $\Delta_K$ in \eqref{deltaOmega} and \eqref{deltaK}, respectively, along with the definitions of the slow-roll parameters $\kappa_U$ and $\sigma_U$ in~\eqref{kappaU}, we obtain
\begin{eqnarray}
\frac{\Omega _{,\varphi}}{\Omega} &=& \frac{f_{,\varphi}}{2f}\ , \\
 \frac{K_{,\varphi}}{2\,K} &=& \frac{E_{,\varphi} }{2\,E}\:  -\: \frac{f_{,\varphi}}{f}\ .
\end{eqnarray}
These two constraining differential equations can be easily solved first for $\Omega$ and then for $K$. In~this~way, we find that 
\begin{equation}
   \label{OmegaKEF}
M^2 \Omega^2\ =\  f\; ,\qquad K\ =\ E/f^2\; .
\end{equation}
However, the solutions for $\Omega$ and $K$, given in~\eqref{OmegaKEF}, single out uniquely the Einstein frame from an arbitrary Jordan frame, provided the mass parameter $M$ is set equal to the reduced Planck mass~$M_P$.

\section{Specific Models}\label{sec:specmod}

In this section we will apply our frame-covariant formalism to a few typical scalar-curvature models of inflation, such as induced gravity inflation and Higgs inflation. In addition, we consider Starobinsky-like $F(R)$ models of inflation, which can be shown to be equivalent to scalar-curvature theories via a Legendre transform, after the introduction of an auxiliary scalar field. In all the examples that we will be considering, we assume that the slow-roll approximation describes well the inflationary dynamics, such that we can use the results presented in Section~\ref{subsec:potsrinf} to derive analytical expressions for all relevant cosmological observables, such as the tensor-to-scalar ratio~$r$, the scalar and tensor spectral indices $n_{\cal R}$ and $n_T$, and their runnings $\alpha_{\cal R}$ and $\alpha_T$.

\subsection{Induced Gravity Inflation}

Induced gravity inflation postulates that the value of the effective Planck mass $M_P$ is exclusively  controlled by the VEV of the inflaton field~$\varphi$. In the Jordan frame, induced gravity inflation is described by a non-minimal coupling $f(\varphi) = \xi \varphi^2$ to the Ricci scalar~$R$, a canonical kinetic term, i.e.~$k(\varphi ) = 1$, and a potential of the form $V(\varphi) = \lambda (\varphi^2 - M_P^2/\xi^2)^2$~\cite{Zee:1978wi,PhysRevD.31.3046}. 

Knowing the explicit forms of the model functions $f(\varphi)$, $k(\varphi )$ and $V(\varphi )$, we may first use them to evaluate the generalized slow-roll parameters~$\epsilon_U$, $\delta_U$, $\kappa_U$ and $\sigma_U$ defined in~\eqref{kappaU}. Then, with the aid of these parameters, we can analytically calculate all relevant inflationary parameters in the slow-roll approximation, as functions of the inflaton field~$\varphi$. In detail, we find
\begin{eqnarray}
   \label{IGIr}
r &=& \frac{128 M_P^4 \xi }{(1+6 \xi  ) \left(M_P^2-\xi  \varphi ^2\right)^2}\ ,\\
   \label{IGInR}
n_{\mathcal{R}} &=& \frac{M_P^4 (1-10 \xi )-2 M_P^2 \xi  (1+ 14 \xi ) \varphi ^2+\xi ^2 (1+ 6 \xi  ) \varphi ^4}{(1+6 \xi ) \left(M_P^2-\xi  \varphi ^2\right)^2} \ ,\\
   \label{IGIaR}
\alpha_{\mathcal{R}} 
&=&
-\frac{128 M_P^4 \xi ^3 \varphi ^2 \left(3 M_P^2+\xi  \varphi ^2\right)}{(1 + 6 \xi  )^2 \left(M_P^2-\xi  \varphi ^2\right)^4} \ ,\\
   \label{IGIaT}
\alpha_{T} 
&=& -\frac{256 M_P^6 \xi ^3 \varphi ^2}{(1+6 \xi)^2 \left(M_P^2-\xi  \varphi ^2\right)^4}\ ,
\end{eqnarray}
with $n_T = -r/8$. In the same slow-roll approximation, the number of e-folds $N$ is found to be  
\begin{equation}
   \label{IGIN}
N\ =\ -\, \frac{(1 + 6 \xi  ) \left[2 M_P^2 \ln \left( {\sqrt{\xi } \varphi  /M_P}\right)+M_P^2-\xi  \varphi ^2\right]}{8\xi M_P^2  }\; .
\end{equation}
Here we used the fact that induced gravity inflation ends at exactly $\varphi = \varphi_{\rm end} = M_P/\sqrt{\xi}$. 

There are two scenarios of gravity induced inflation: (i)~the scenario of small-field inflation, in which the inflaton starts at small values, in which $\xi \varphi^2 \ll M_P^2$, and (ii)~the scenario of standard chaotic large-field inflation, in which $\xi \varphi^2 \gg M_P^2$. Ideally, we wish to express quantities  in terms of the number~$N$ of e-folds. To do this, we must invert the relation $N=N(\varphi)$ to $\varphi = \varphi (N)$, in order to substitute the latter into the inflationary observables. However, this proves to be challenging for most models. Therefore, our strategy will be to expand $N(\varphi)$ (about zero for small-field inflation, and about infinity for large-field inflation), truncate the series to lowest order, and then invert this truncated relation, \emph{before} substituting it into the expressions for the inflationary observables. This will help us to make contact with already established results in the literature, while simultaneously allowing for more accurate predictions to be extracted simply by including more terms in the series expansion.

\subsubsection{Small-Field Inflation}

For small-field (SF) inflation, a good approximation is obtained, if the cosmological observables listed in~\eqref{IGIr}--\eqref{IGIaT} and the number of e-folds~$N$ in~\eqref{IGIN} are expanded about $\varphi = 0$, thus assuming that the horizon exit happened long before the end of inflation. Thus, for SF values of~$\varphi$,  the number of e-folds $N$  becomes
\begin{equation}
    \label{efoldsIGI}
N\ =\ -\frac{1+6 \xi  }{8 \xi }\left[  \ln \left(\frac{\xi\varphi^2     }{M_P^2}\right)\:
+\: 1\right]\, .
\end{equation}
From this last expression, we  see that a large number of e-folds $N$ corresponds to \emph{small} values of~$\varphi$. 

We may now invert the relation $N = N(\varphi )$ given in~\eqref{efoldsIGI}, i.e.~as $\varphi = \varphi (N)$, so as to write the cosmological observables in terms of~$N$. Hence, in the SF approximation for~$\varphi$, we get
\begin{eqnarray}
\label{IGIobsSF}
r_{\text{SF}}  &=& \frac{128 \xi  e^{2\beta_\xi N +2}}{(1+ 6 \xi) 
\left(e^{\beta_\xi N +1}-1\right)^2}\ ,\nonumber\\
 n_{\mathcal{R},\text{SF}} &=&  \frac{ 1+6 \xi  -2 (1+14 \xi  ) e^{\beta_\xi N +1}+(1-10 \xi ) e^{2\beta_\xi N +2}
}{(1+6 \xi  ) \left(e^{\beta_\xi N +1}-1\right)^2}\ ,\nonumber\\
\alpha_{\mathcal{R},\text{SF}}  &=& -\frac{128 \xi ^2 e^{2\beta_\xi N+2} \left(3 e^{\beta_\xi N +1}+1\right)}{(1+ 6 \xi)^2 \left(e^{\beta_\xi N+1}-1\right)^4}\ ,\nonumber\\
\alpha_{T,\text{SF}}  &=& -\frac{256 \xi ^2 e^{3\beta_\xi N +3}}{(1+ 6 \xi)^2 \left(e^{\beta_\xi N +1}-1\right)^4}\ ,
\end{eqnarray}
with $\beta_\xi = 8\xi/(1+6\xi)$.
In particular, for a large number $N$ of e-folds, we find
\begin{equation}
\label{IGIobsSFlargeN}
r_{\text{SF}}\ \simeq\
\frac{128 \xi }{1+6 \xi  }\ ,\qquad
 n_{\mathcal{R},\text{SF}}\ \simeq\
 1\: -\: \frac{16 \xi }{1+6 \xi  }\ .
\end{equation}
Consequently, the tensor-to-scalar ratio~$r$ and the scalar spectral index~$n_{\cal R}$ are not sensitive to $N$ in the gravity induced scenario of  SF inflation. Expressions similar to \eqref{IGIobsSFlargeN} have been reported in the literature~\cite{Kaiser:1994vs}, all of which are approximations of \eqref{IGIobsSF} for a large number~$N$ of e-folds.

\subsubsection{Large-Field Inflation}

For the case of the gravity induced scenario of large-field (LF) inflation,  we expand the 
analytical expressions~\eqref{IGIr}--\eqref{IGIN}  of all relevant inflationary quantities given  in terms of $\varphi$ about infinity. In this LF limit, the number of e-folds simplifies to
\begin{equation}
N\ =\ \frac{(1+ 6 \xi ) \varphi ^2}{8 M_P^2}\ .
\end{equation}
Substituting this expression in \eqref{IGIr}--\eqref{IGIaT}, the inflationary observables in the same LF limit  become
\begin{eqnarray}
\label{IGIobsLF}
r_{\text{LF} }   &=&  \frac{128\, \xi\,  (1+6 \xi)}{[(8 N-6) \xi -1]^2}\ ,\nonumber\\
 n_{\mathcal{R},\text{LF} } &=& \frac{4 \left(16 N^2-56 N-15\right) \xi ^2-4 (4 N+1) \xi +1}{[(8 N-6) \xi -1]^2}\ ,\nonumber\\
\alpha_{\mathcal{R}, \text{LF} }   &=&  -\frac{1024 N \xi ^3 (2 (4 N+9) \xi +3)}{[(6-8 N) \xi +1]^4}\ ,\nonumber\\
\alpha_{T,\text{LF} }   &=&  -\, \frac{2048 N \xi ^3 (1+6 \xi)}{[(6-8 N) \xi +1]^4}\ .
\end{eqnarray}
Upon expanding for a large number $N$ of e-folds, the above expressions simplify to    
\begin{eqnarray}
r_{\text{LF} }   &=& \left(\frac{3}{4} + \frac{1}{8 \xi }\right)\frac{1}{N^2}\ 
+\ {\cal O}\left(\frac{1}{N^3}\right)\; ,\nonumber\\[3mm]
 n_{\mathcal{R},\text{LF} }& =& 1\ -\ \frac{2}{N}\ 
+\ {\cal O}\left(\frac{1}{N^2}\right)\;,
\nonumber\\[3mm]
\alpha_{R,\text{LF}}   &=& - \frac{2}{N^2}\ 
 +\ {\cal O}\left(\frac{1}{N^3}\right)\;,
\nonumber\\[3mm]
\alpha_{T,\text{LF}} &=& -\left( 3 + \frac{1}{2 \xi }\right)\frac{1}{N^3}\ 
+\ {\cal O}\left(\frac{1}{N^4}\right)\; .
\end{eqnarray}

Finally, it is interesting to evaluate the admissible value of $\lambda$ by the normalization of the power spectrum,
\begin{equation}
\Delta_\mathcal{R}\ =\ \frac{\lambda\,  (1+6 \xi  ) 
\left( M^2-\xi  \varphi ^2\right)^4}{32 M^4 \xi ^5 \varphi ^4}\ .
\end{equation}
In terms of $N$, this is given by
\begin{equation}
\Delta_\mathcal{R}\ =\ \frac{\lambda\,  [1+(6-8 N) \xi  ]^4}{  2048 N^2 \xi ^5 (1+6 \xi)}\ .
\end{equation}
The power spectrum is normalised via \cite{Friedman:2006zt}
\begin{equation}
   \label{COBEnorm}
\Delta_{\cal R}\ =\ \frac{1}{3}\: \frac{U}{\epsilon_U + \kappa_U}\  =\ \frac{(0.027)^4}{3}\ ,
\end{equation}
where $U(\varphi ) \equiv V(\varphi )/f^2(\varphi )$ [cf.~\eqref{Udef}], and 
$\epsilon_U$ and $\kappa_U$ are given in~\eqref{kappaU}. Hence, the value of the quartic coupling~$\lambda$ may be estimated in terms of the non-minimal coupling~$\xi$ as follows:
\begin{equation}
 \lambda\ \approx\  (0.027)^4\times\; \frac{\xi  (1+6 \xi )}{2N^2}\ ,
\end{equation}
where $N \approx 60$ is the scale at which the largest cosmological scales have presently re-entered the horizon. We should reiterate here that, to leading order in $1/N$, our analytical predictions for the cosmological observables $\Delta_{\cal R}$, $r$ and $n_{\cal R}$  reproduce the results known from the literature~\cite{Kaiser:1994vs} for both the scenarios of SF and LF induced gravity inflation. Within our frame-covariant formalism, however, the full frame-invariant expressions for all inflationary quantities can be computed to arbitrarily high order in $1/N$, simply by including higher-order terms in the expansion for $N(\varphi)$.

\subsection{Higgs Inflation}

This scenario is based on the radical suggestion~\cite{Salopek:1988qh,bezrukov08} that the inflaton field~$\varphi$ {\em is} the Standard Model~(SM) Higgs boson observed at the CERN Large Hadron Collider (LHC). In order to make such a scenario phenomenologically viable, however, a sizable non-minimal coupling $\xi$ of the Higgs field~$\varphi$ to the curvature $R$ is required, which may be partly attributed to renormalization-group running effects, even within the SM in curved space~\cite{Herranen:2014cua}.

In the Jordan frame, the model of Higgs inflation can be fully described by the non-minimal coupling function $f(\varphi) = M_P^2 + \xi \varphi^2$, a canonical kinetic term for the inflaton ($k(\varphi ) = 1$),  and the SM potential: $V(\varphi) = \lambda (\varphi^2 - v^2)^2$, where $v$ is the VEV of the Higgs boson. As before, we apply our frame-covariant approach of inflation  in the slow-roll approximation to analytically compute the cosmological observables, i.e.
\begin{eqnarray}
r &=& \frac{128 M_P^4}{\varphi ^2 \left[M_P^2+\xi  (1+6 \xi) \varphi ^2\right]}\ ,\nonumber\\
n_\mathcal{R} &=&
\frac{\xi ^2 (1+6 \xi  )^2 \varphi ^6 
-2 M_P^2 \xi  \left(48 \xi ^2+2 \xi -1\right) \varphi ^4 
+M_P^4 \left(1 -40 \xi -192 \xi ^2 \right) \varphi ^2
-24 M_P^6}{\varphi ^2 \left[M_P^2+\xi  (1+ 6 \xi) \varphi ^2\right]^2}\ ,\nonumber\\
\alpha_\mathcal{R} &=& -\frac{64 M_P^4 \left(M_P^2+\xi  \varphi ^2\right) \left[3 M_P^6+9 M_P^4 \xi  (1+6 \xi) \varphi ^2+8 M_P^2 \xi ^2 (1+6 \xi)^2 \varphi ^4+2 \xi ^3 (1+6 \xi)^2 \varphi ^6\right]}{\varphi ^4 \left[M_P^2+\xi  (1+6 \xi) \varphi ^2\right]^4}\ ,\nonumber\\
\label{higgsstensrun}
\alpha_T &=& -\frac{128 M_P^6 \left(M_P^2+\xi  \varphi ^2\right) \left[M_P^2+2 \xi  (1+6 \xi) \varphi ^2\right]}{\varphi ^4 \left[M_P^2+\xi  (1+6 \xi) \varphi ^2\right]^3}\ ,
\end{eqnarray}
with $n_T = -r/8$. Assuming that the field value~$\varphi$ at horizon exit is much larger than that at the end of inflation, i.e.~$\varphi\gg \varphi_\text{end}$, the number $N$ of e-folds reads
\begin{equation}\label{efoldshiggs}
N\ =\  \frac{(1+6 \xi ) \varphi ^2}{8M_P^2} +\frac{6}{8}\ln \left(\frac{M_P^2}{M_P^2+\xi  \varphi ^2}\right)\; .
\end{equation}
Under the assumption $\xi\varphi^2 \gg M_P^2$, after inverting \eqref{efoldshiggs}, we obtain to leading order,
\begin{equation}
\varphi\ =\ \left(\frac{8M_P^2 N}{1+6 \xi} \right)^{1/2}.
\end{equation}
Substituting this last expression in~\eqref{higgsstensrun} leads to
\begin{eqnarray}
    \label{higgsobs}
r &=& \frac{ 16 (1+6 \xi)}{8 \xi  N^2+N}\nonumber\\
&=& \left(12 + \frac{2}{\xi}\right) \frac{1}{N^2}\ 
+\  
{\cal O}\left(\frac{1}{N^3}\right)\;, \nonumber\\[5mm]
n_\mathcal{R}
&=&
\frac{64 \xi ^2 N^3  + \left(1 -40 \xi -192 \xi ^2 \right) N-16 \xi  (8 \xi -1) N^2- 3 (1+6 \xi)}{N (1+8 \xi  N)^2}\hspace{3cm}{}\nonumber\\
&=& 
1\ -\ \frac{2}{N}\ 
+\ 
{\cal O}\left(\frac{1}{N^2}\right)\; ,\nonumber
\end{eqnarray}
\begin{eqnarray}
\alpha_\mathcal{R} &=&
-\frac{1}{N^2 (1 + 8 \xi N )^4}\; \bigg[ 2048 \xi ^3 N^2 \left(4 N^2+15 N+9\right) \xi ^4+32 N \left(160 N^2+300 N+81\right) \nonumber\\
&&+\; 4 \left(272 N^2+252 N+27\right) \xi ^2+12 (8 N+3) \xi  + 3\,\bigg]\nonumber\\
&=& -\frac{2}{N^2}\ 
+\ 
{\cal O}\left(\frac{1}{N^3}\right)\;,\nonumber \\[5mm]
\alpha_T &=& -\frac{2 (1 + 6 \xi) \left(32 N (4 N+3) \xi^2
+6 (4 N+1) \xi +1\right)}{N^2 (8 \xi N +1)^3} \nonumber\\
&=& -\,\left( 3 + \frac{1}{2 \xi }\right)\frac{1}{N^3}\ 
+\ 
{\cal O}\left(\frac{1}{N^4}\right)\; .
\end{eqnarray} 
In the above, we also quote approximate results for large values of e-folds~$N$, assuming that 
$\xi \stackrel{>}{{}_\sim} 1$. 

We may now estimate the size of $\xi$, using the normalization of the dimensionless power spectrum~$\Delta_{{\cal R}}$. The power spectrum in terms of the inflaton field~$\varphi$ is given by
\begin{equation}
\Delta_\mathcal{R}\ =\ \frac{\lambda  \varphi ^6 \left[M^2+\xi  (1 + 6 \xi) \varphi ^2\right]}{32 M_P^4 \left(M_P^2+\xi  \varphi ^2\right)^2}\ ,
\end{equation}
which may be translated into the number~$N$ of e-folds as
\begin{align}
\Delta_\mathcal{R}\ =\ \frac{16 \lambda  N^3 (8 \xi  N+1)}{ (1+6 \xi  ) [  (8 N+6)\xi+1]^2}\ .
\end{align}
Setting $\lambda = 0.129$ as the value for the quartic coupling (corresponding to a SM Higgs-boson mass of 125~GeV) and $N=60$ as the nominal number of e-folds for
the horizon exit, we may match $\Delta_\mathcal{R}$ with the normalization~\eqref{COBEnorm}
to deduce the known result~\cite{bezrukov08}:
\begin{align}
 \xi\ =\ \frac{  N }{ \sqrt{3} }\; \frac{\sqrt{\lambda}}{(0.027)^2  }\ \approx\ 17,000\; .
\end{align}
Note that our method for deriving observables does not require transforming to the Einstein frame, in which finding a closed-form expression of the potential is not possible. Hence, our frame-covariant approach leads to more accurate results which still agree to leading order in $1/N$ with those reported in the literature~\cite{bezrukov08,Salopek:1988qh}.

\subsection{$F(R)$ Models}
\label{FRmod}

An interesting class of possible inflationary scenarios emerges, when the universe self-accelerates without the direct presence of a scalar field~\cite{Starobinsky:1980te}. A wide range of such models is encoded in $F(R)$ theories, which are described by the following action:
\begin{equation}
   \label{SFR}
S[g_{\mu\nu},F]\ =\ -\int d^4 x\, \sqrt{-g} \: \frac{F(R)}{2}\ .
\end{equation}
These theories may be recast in a form equivalent to the scalar-curvature theories by introducing an auxiliary field~$\Phi$:
\begin{equation}
    \label{actionR}
S[g_{\mu\nu},\Phi ] \ =\ - \int d^4 x \sqrt{-g} \, \frac{1}{2}\Big[ F(\Phi)\: +\: F(\Phi)_{,\Phi} (R-\Phi) \Big]\,.
\end{equation}
It is not difficult to check that the equation of motion for $\Phi$, $\delta S/\delta\Phi = 0$, implies $\Phi = R$, provided $F(\Phi)_{,\Phi\Phi}$ does not vanish in the domain of interest. Consequently, the action in~\eqref{actionR} is equivalent to the original action of $F(R)$ theories given in~\eqref{SFR}.

We may now introduce another field $\varphi$, such that
\begin{align}\label{Fphi}
\varphi\ =\ F(\Phi)_{,\Phi}\; ,
\end{align}
which will play the role of the inflaton. To this end, we write~\eqref{actionR} as
\begin{equation}
   \label{actionRphi}
S[g_{\mu\nu},\varphi]\ =\ \int d^4 x \sqrt{-g} \: \bigg[ -\frac{1}{2}\varphi\,  R\:  +\: V(\varphi)\,\bigg]\,,
\end{equation}
where $V(\varphi)$ is given by
\begin{equation} 
   \label{VFR}
V(\varphi)\ =\ \frac{1}{2}\,\varphi\,\Phi(\varphi)\: -\: \frac{1}{2}\,F\big(\Phi(\varphi)\big)\; .
\end{equation}
Here, the expression for $\Phi = \Phi(\varphi)$ comes from inverting the functional relation  in~\eqref{Fphi}. This action is equivalent to a special class of scalar-curvature theories, termed \emph{Brans-Dicke models}~\cite{Mathiazhagan:1984vi}, with the additional constraint: $k(\varphi)=0$, i.e.~the absence of an inflaton kinetic term in the considered Jordan frame.

We will now present some typical results that can be obtained for a simple class of $F(R)$ theories, within our frame-covariant formalism of inflation. We consider a modified version of
the  Starobinsky model \cite{Starobinsky:1980te} that still offers analytic predictions. In this version of Starobinsky-like inflation, the function $F(R)$ assumes the form
\begin{equation}
   \label{FRstar}
F(R)\ =\ \alpha R\: +\: \beta_n R^n \; ,
\end{equation}
where $\alpha$ and $\beta_n$ are arbitrary parameters and $n\ge 2$.  The usual procedure would be to perform a conformal transformation of the action \eqref{actionRphi} to go to the Einstein frame~\cite{Sotiriou:2008rp}. Within our frame-covariant approach, however, this intermediate computational step becomes unnecessary. Instead, we may simply use the functional form of $F(R)$ to arrive at an expression for the potential $V(\varphi)$, as given in~\eqref{VFR}, and derive predictions for the cosmological observables by considering the equivalent scalar-curvature theory in the Jordan frame.

Given the form of $F(R)$ in~\eqref{FRstar}, \eqref{Fphi} yields
\begin{equation}
   \label{PhiVarphi}
\varphi(\Phi)\ =\ \alpha\: +\:  \beta_n n \Phi^{n-1}\; ,
\end{equation}
which is easily inverted to
\begin{equation}
\Phi(\varphi)\ =\ \left(\frac{\varphi-\alpha}{ \beta_n n} \right)^{1/(n-1)}\; .
\end{equation}
The potential thus becomes
\begin{equation} 
    \label{VFRphi}
 V(\varphi)\ =\   \frac{n-1}{2 }\beta_n \left(\frac{\varphi -\alpha}{ \beta_n n}\right)^{n/({n-1)} }\;.
\end{equation}

For this class of Starobinsky-like theories, the model functions are: $f(\varphi ) = \varphi$, $k (\varphi ) = 0$, and $V(\varphi )$ is given by~\eqref{VFRphi}. Applying the results of our frame-covariant formalism presented in~Section~\ref{subsec:potsrinf}, the following analytic expressions for the cosmological parameters are obtained:
\begin{eqnarray}
   \label{FRr}
r &=& \frac{16}{3} \frac{  [(n-2) \varphi -2 \alpha (n-1)]^2}{ (n-1)^2 (\varphi- \alpha )^2}\ ,\\
   \label{FRnR}
n_\mathcal{R} &=& \frac{\left(n^2+2 n-5\right) \varphi ^2
-2 \alpha \left(n^2+4 n-5\right) \varphi -5 \alpha^2 (n-1)^2}{3 (n-1)^2 (\varphi - \alpha)^2}\ ,\\
   \label{FRaR}
\alpha_\mathcal{R} &=&
 \frac{8 \alpha n \varphi  (3 \alpha (n-1)+\varphi ) [(n-2) \varphi -2 \alpha (n-1)]}{9 (n-1)^3 (\varphi-\alpha )^4}\ ,\\
   \label{FRaT}
\alpha_T &=& -\frac{8 \alpha n \varphi  [(n-2) \varphi -2 \alpha (n-1)]^2}{9 (n-1)^3 
(\varphi - \alpha)^4}\ ,
\end{eqnarray}
with $n_T = -r/8$. As before, the inflaton field value $\varphi$ must be evaluated at the point of horizon crossing. We note that generically, $\varphi$ starts small during inflation and gets even smaller as the number of e-folds increases. Hence, we calculate the number of e-folds~$N$ by expanding $\varphi$ about $\varphi_\text{end}$ to lowest order:
\begin{equation} 
   \label{NphiFR} 
N\ =\ -\, \frac{3}{2}\:\frac{ (n-1) (\varphi -\varphi_\text{end})   (\varphi_\text{end}-\alpha) }{  \varphi_\text{end} [(n-2) \varphi_\text{end}-2 \alpha (n-1)]}\ .
\end{equation}
At the end of inflation, we expect that $F(R) = M_P^2 R$, i.e.
\begin{align} 
F(R_\text{end})\ =\ \alpha R_\text{end} + \beta_n R_\text{end}^n\ =\ M_P^2 R_\text{end}\;.
\end{align}
Since $\Phi = R$, we find
\begin{align} 
\Phi_\text{end} \ =\ \left( \frac{M_P^2  - \alpha}{\beta_n} \right)^{1/(n-1)}\,.
\end{align}
From \eqref{PhiVarphi}, it is then possible to calculate $\varphi_\text{end}$ as
\begin{equation} 
\varphi_\text{end}\ =\  nM_P^2  - (n-1)\alpha\; .
\end{equation}
As a consequence, the number of e-folds $N$ in~\eqref{NphiFR} becomes
\begin{align} 
  \label{NeFR}
N\ =\ \frac{3 (n-1) \left(\alpha-M_P^2\right) \left[\alpha (n-1) -n M_P^2  +\varphi \right]}{2 \left[ M_P^2 (n-2) - \alpha(n-1) \right] \left[ M_P^2 n -\alpha(n-1) \right]}\ .
\end{align}
Solving~\eqref{NeFR} for $\varphi$, substituting its expression into~\eqref{FRr}--\eqref{FRaT}, and expanding the latter for large~$N$ to order $1/N$, we derive the following approximate analytic expressions for the cosmo\-logical observables: \begin{eqnarray} 
   \label{FRapprox}
r &\approx& \frac{16 (n-2)^2}{3 (n-1)^2}\ -\ \frac{16 \alpha (n-2) n \left(\alpha-M_P^2\right)}{  (n-1)  \left[   M_P^2 (n-2) -\alpha(n-1)\right] \left[ M_P^2 n-\alpha(n-1)  \right]}\, \frac{1}{N}\ ,\nonumber\\
 n_\mathcal{R} &\approx& \frac{n^2+2 n-5}{3 (n-1)^2} \ -\ \frac{2 \alpha n \left(\alpha-M_P^2\right)}{  (n-1) \left[\alpha^2 (n-1)^2-2 \alpha M_P^2 (n-1)^2+M_P^4  (n-2) \right]}\, \frac{1}{N}\ , \nonumber\\
\alpha_\mathcal{R} &\approx& \, \frac{4 \alpha (n-2) n \left(\alpha-M_P^2\right)}{3   (n-1)^2 \left[  M_P^2 (n-2)-\alpha(n-1) \right] \left[  M_P^2 n-\alpha(n-1)\right]}\, \frac{1}{N}\ ,\nonumber\\
\alpha_T &\approx&  -\frac{4 \alpha  (n-2)^2 n \left(\alpha-M_P^2  \right)}{3   (n-1)^2 \left[ M_P^2 (n-2)  -\alpha  (n-1)\right]\left[  M_P^2 n-\alpha  (n-1)\right] }\, \frac{1}{N}\ .
\end{eqnarray}
We observe that $\beta_n$ does not enter the expressions for the observables. In fact, all inflationary observables are independent of $\beta_n$, to {\em all} orders in $1/N$. Instead, we see that there is strong dependence on the power $n$ of $R^n$ in~\eqref{FRstar}, and for $\alpha = M_P^2$, the expressions listed in~\eqref{FRapprox} become independent of the number of e-folds~$N$ through order $1/N$. Finally, we note that, for $\alpha \neq M_P^2$, the runnings of the spectral indices $\alpha_\mathcal{R}$ and $\alpha_T$ start at order $1/N$, and so they turn out to be at least one order of magnitude larger than those found in the models of induced gravity and Higgs inflation.

\section{Beyond the Tree-Level Approximation}\label{sec:BCA}

In the process of developing a frame-covariant formalism of inflation, we have assumed that the inflaton and metric perturbations are quantized fields. By using the equations of motion to derive expressions for the mode functions and thus the correlation functions, we have been calculating 
all relevant inflationary observables at the tree level only. However, higher order radiative corrections may induce a non-negligible correction to the inflationary observables. At this time, the question whether these quantum corrections to cosmological observables are frame-invariant has not yet been resolved \cite{George:2013iia,Kamenshchik:2014waa,Domenech:2015qoa}. It has been suggested \cite{Steinwachs:2013tr,Kamenshchik:2014waa,Moss:2014} that the Vilkovisky--DeWitt formalism \cite{Vilkovisky:1984st, DeWitt} could be used to solve the frame problem beyond the tree level approximation. In this section, we will outline how to extend the frame invariance of the action \eqref{actiontransprop} to the \textit{effective action}, which incorporates the aforementioned corrections, through the use of the Vilkovisky--DeWitt formalism. We will explicitly demonstrate this invariance at the one-loop level.

In order to simplify the discussion, we shall make two assumptions: (i) the inflaton field does not couple to other matter fields, even though their inclusion will be straightforward in the present formalism, and (ii) the radiative corrections coming from the quantized metric perturbation are negligible in comparison to the quantum inflaton corrections. Assumption (ii) will be sufficient in most cases, as quantum gravitational corrections will be ${\cal O}(1/M_P^2)$ in general, and as such they can be ignored in comparison to the quantum inflaton corrections.

The quantum-corrected inflaton equation of motion and Einstein field equations are given by
\begin{equation}
\frac{\delta \Gamma}{\delta \varphi(x)}\: =\: 0 \;, \qquad 
\frac{\delta \Gamma}{\delta g_{\mu \nu}(x)}\: =\: 0\;,
\end{equation}
where $\delta/\delta\varphi(x)$ and $\delta/\delta g_{\mu \nu}(x)$ are functional derivatives with respect to the fields $\varphi(x)$ and $g_{\mu \nu}(x)$ respectively, and $\Gamma[g_{\mu \nu}, \varphi] \equiv \Gamma[g_{\mu \nu}, \varphi, f(\varphi), k(\varphi), V(\varphi)]$ is the effective action which is determined through the functional integro-differential equation
\begin{equation}
    \label{eq:effacteqn}
\exp\bigg(\frac{i}{\hbar} \Gamma[g_{\mu \nu}, \varphi]\bigg)\: =\: \int \mathcal{D} \varphi^Q \mathcal{M}[\varphi^Q]\: \exp\bigg(\frac{i}{\hbar}\bigg[S[g_{\mu \nu}, \varphi^Q] - \int d^4 x\,(\varphi - \varphi^Q) \frac{\delta \Gamma [g_{\mu \nu}, \varphi]}{\delta \varphi}\bigg]\bigg) \; , 
\end{equation}
where $\mathcal{D}\varphi^Q \,\mathcal{M}[\varphi^Q]$ is the path integral measure and $S[g_{\mu \nu}, \varphi]$ is the action defined in \eqref{actionJ}. 

To obtain an expression for the effective action, we shall solve equation \eqref{eq:effacteqn} perturbatively in $\hbar$. To make this process simpler, we may make a field transformation of the quantum field $\varphi^Q$ to $\varphi^{\prime Q}$, given by $\varphi^{Q} = \varphi + \hbar^{\frac{1}{2}} \varphi^{\prime Q}$. We then expand the effective action in powers of $\hbar$:
\begin{equation}
\Gamma[g_{\mu \nu}, \varphi]\ =\ \sum_{n = 0}^\infty \hbar^n \Gamma_n[g_{\mu \nu}, \varphi]\;.
\end{equation}
For simplicity, we shall only compute $\Gamma$ to ${\cal O}(\hbar)$. We find
\begin{align}
\Gamma_0[g_{\mu \nu},\varphi]\ &=\ S[g_{\mu \nu},\varphi]\,, \\
 \label{oneloopnaive}
\Gamma_1[g_{\mu \nu},\varphi]\  &=\ \ln \mathcal{M}[\varphi]\: -\: \frac{1}{2} \ln  \det \left(\frac{\delta^2 S[g_{\mu \nu},\varphi] }{\delta \varphi(x) \delta \varphi(y)} \right) \; .
\end{align}
Now that we know the explicit expression for $\Gamma_1[g_{\mu \nu},\varphi]$, we shall examine how it transforms under inflaton reparameterizations and conformal transformations.

Let us first consider inflaton reparameterizations within the one-loop effective action $\Gamma_1[g_{\mu \nu},\varphi]$. For this discussion, we shall denote $S[\varphi, k(\varphi)] \equiv S[g_{\mu\nu},\varphi, f(\varphi), k(\varphi), V(\varphi)]$ and $ \Gamma_1[\varphi, k(\varphi)] \equiv \Gamma_1 [g_{\mu\nu},\varphi,f(\varphi), k(\varphi), V(\varphi)]$ for brevity, since only $\varphi$ and $k(\varphi)$ are affected by the inflaton repara\-meterizations. Under the transformations \eqref{Kphi}, we may write $\Gamma_1$ as
\begin{align}\label{effacttrans}
\Gamma_1 [\tilde\varphi, \tilde k(\tilde\varphi)]\ =\ \ln \mathcal{ \widetilde M}[ \tilde \varphi]   \: -\: \frac{1}{2} \ln \det \left( \frac{\delta^2 S[\tilde\varphi, \tilde k(\tilde\varphi)] }{\delta \tilde \varphi(x) \delta \tilde\varphi(y)  } \right) \,.
\end{align}
We wish to relate $\Gamma_1[\varphi, k(\varphi)]$ to $\Gamma_1 [\tilde \varphi, \tilde k (\tilde \varphi)]$. We first examine how the first functional derivative of the action transforms. We obtain
\begin{align} 
\frac{\delta S  [\varphi, k(\varphi)]}{\delta \varphi(x)}
\ &=\  
\sqrt{-g} \left(-\frac{f(\varphi)_{,\varphi}}{2}R - \frac{k(\varphi)_{,\varphi}}{2} (\partial  \varphi)^2
- k(\varphi) \partial^2  \varphi
- V(\varphi)_{,\varphi}\right)\,,
\nonumber\\
\label{actionjac}
\frac{\delta  S [\tilde\varphi, \tilde k(\tilde\varphi)]}{\delta \tilde \varphi(x)}\ &=\ 
K_x^{-1/2} \,\frac{\delta S  [\varphi, k(\varphi)]}{\delta \varphi(x)}\, ,
\end{align}
where we use \eqref{Imodelparamdef} with $\Omega = 1$ and have denoted $K_x \equiv K(\varphi(x))$ for brevity. We may now proceed to determine the transformation of the second functional derivative of the action which we find to be
\begin{align}
    \label{seconddertrans}
\frac{\delta^2 S  [\tilde\varphi, \tilde k(\tilde\varphi)]}{\delta \tilde \varphi(x)\delta \tilde\varphi(y)} 
\ &=\ 
  K^{-1/2}_x K^{-1/2}_y  \left[\frac{\delta^2 S  [\varphi, k(\varphi)]}{\delta \varphi(x)   \delta \varphi(y)}\:  -\: \frac{  K_x^{- 1/2}}{2}   (\ln K_x)_{,\varphi}   \frac{\delta S  [\varphi, k(\varphi)]}{\delta \varphi(x)}\delta(x-y)\right].
\end{align}
Now that we have found the transformation of the second term in~\eqref{effacttrans}, let us consider how the function $\mathcal{M}[\varphi]$ transforms under inflaton reparametrizations. We will require that the path integral measure must remain invariant under inflaton parameterizations:
\begin{equation}
\mathcal{D}\varphi^Q \,\mathcal{M}[\varphi^Q]\ =\ \mathcal{D} \tilde{\varphi}^Q\, \mathcal{\widetilde{M}}[\varphi^Q(\tilde \varphi^Q)]
\end{equation}
and hence this defines the transformation of $\mathcal{M}$ as
\begin{align}\label{measuretrans}
\mathcal{ \widetilde M}[ \tilde \varphi]\ \equiv\ \det \big(K^{-1/2}_x \delta(x-y)\big) \mathcal{M}[\varphi ( \tilde \varphi)].
\end{align} 
Consequently, due to the combination of the transformations \eqref{seconddertrans} and \eqref{measuretrans}, the one-loop effective action $\Gamma_1$ is \textit{not} invariant under inflaton reparametrizations:
\begin{equation}
\Gamma_1[\tilde \varphi, \tilde k(\tilde \varphi)]\ \neq\ \Gamma_1[\varphi, k(\varphi)] \, .
\end{equation}
This result is also consistent with the one observed in~\cite{Steinwachs:2013tr}, by an explicit computation. As shown in~\eqref{seconddertrans}, the primary source of the frame-dependence is the presence of the functional derivatives with respect to $\varphi(x)$. It was the idea of Vilkovisky \cite{Vilkovisky:1984st} to extend the effective action such that it remains invariant under field reparameterizations. With subsequent developments by DeWitt \cite{DeWitt}, the combined work is now known as the \textit{Vilkovisky--DeWitt formalism}. The essential observation is the following. If the second term in \eqref{oneloopnaive} transforms \textit{covariantly}, then the one-loop effective action would remain invariant under inflaton reparametrizations.

Let us briefly describe Vilkovisky's idea. Suppose we identify the field $\varphi(x)$ at each spacetime point $x$ with a coordinate on a manifold. We shall call this manifold the \textit{field space}. Given this identification, it is possible to extend the notion of functional derivatives to \emph{covariant functional derivatives}. Denoting the covariant functional derivative by $D/D\varphi(x)$, we have
\begin{align}
\frac{D^2 S}{D \varphi(x)D \varphi(y)}\ \equiv\ 
\frac{\delta^2 S }{\delta \varphi(x) \delta \varphi(y)}\:
 -\: \Gamma^z_{xy} \frac{\delta S}{ \delta \varphi(z)}\; ,
\end{align}
where $\Gamma^z_{xy}$ is the connection and we use the Einstein--DeWitt convention in which repeated spacetime coordinates are integrated over all spacetime. In this instance, the connection $\Gamma^z_{xy}$ transforms in a way that ensures that the second covariant functional derivative transforms covariantly under inflaton reparametrizations. To determine the transformation of the connection on the field space, we require that it should transform to cancel the second term inside the brackets in \eqref{seconddertrans}. This requirement leads to the transformation
\begin{align}\label{conntrans}
\widetilde \Gamma^z_{xy}\ =\ K^{1/2}_z K^{-1/2}_x K^{-1/2}_y \left[\Gamma^z_{xy}\: 
-\: \frac{1}{2} ( \ln K_x)_{,\varphi}\delta(x-y) \delta(y-z)\right] .
\end{align}
This ensures that the double covariant functional derivative transforms as
\begin{align}\label{acttenstrans}
\frac{\widetilde D^2 S[\tilde\varphi, \tilde k(\tilde\varphi)] }{\widetilde D \varphi(x) \widetilde D \varphi(y)}\ =\ K^{-1/2}_x K^{-1/2}_y\frac{D^2 S}{D \varphi(x)D \varphi(y)}\ .
\end{align}
Given this transformation property, let us now define a new effective action by replacing the functional derivatives with covariant functional derivatives in the one-loop effective action \eqref{oneloopnaive}:
\begin{align}
\Gamma_1^{\rm VD}[\varphi, k(\varphi)]\ \equiv\ \ln \mathcal{M}[\varphi]\: -\: \frac{1}{2} \ln  \det \left(\frac{D^2 S[\varphi, k(\varphi)] }{D \varphi(x) D \varphi(y)} \right) .
\end{align}
This is known as the one-loop \textit{Vilkovisky--DeWitt effective action}, which can be rigorously derived by generalizing the source term coupled to the fields, such that the linear expression
$\varphi - \varphi^Q$ in~\eqref{eq:effacteqn} is replaced with a function $\sigma (\varphi, \varphi^Q)$ endowed with specific properties~\cite{Vilkovisky:1984st}. Now, if we make an inflaton reparameterization using \eqref{acttenstrans} and \eqref{measuretrans}, we find that  $\Gamma_1^{\rm VD}[\tilde\varphi, \tilde k(\tilde\varphi)]$ transforms as 
\begin{equation} 
\Gamma_1^{\rm VD} [\tilde\varphi, \tilde k(\tilde\varphi)] 
\ =\   \ln \mathcal{   M}[   \varphi] \:   
-\: \frac{1}{2} \ln  \det \left( \frac{D^2 S[  k(  \varphi),  \varphi] }{D   \varphi(x) D  \varphi(y)  } \right)\ =\ \Gamma_1^{\rm VD} [\varphi, k(\varphi)]\; .
\end{equation}
Thus, replacing functional derivatives with their covariant counterparts in the usual definition of the one-loop effective action ensures that the one-loop effective action is unaffected by inflaton reparameterizations. 

Let us now discuss the form of the measure functional $\mathcal{M}[\varphi]$ and the connection $\Gamma^z_{xy}$. Taking the analogous case of differential geometry as an example, one is able to obtain the invariant integral measure and the affine connection in terms of the metric tensor of the field space. From this case, we may construct $\mathcal{M}[\varphi]$ and $\Gamma^z_{xy}$ by taking inspiration from differential geometry in terms of some metric $\mathcal{G}_{xy}$ for the field space. These expressions then take the form
\begin{align}
\mathcal{M}[\varphi]\ &\equiv\ \sqrt{\det \mathcal{G}_{xy}}\, ,\\ 
\Gamma^{z}_{xy}\  &\equiv\ \frac{1}{2}\mathcal{G}^{zw}\bigg(\frac{\delta \mathcal{G}_{wx}}{\delta \varphi(y)} + \frac{\delta \mathcal{G}_{wy}}{\delta \varphi(x)} - \frac{\delta \mathcal{G}_{xy}}{\delta \varphi(w)}\bigg)\,,
\end{align}
where $\mathcal{G}^{zw}$ is the inverse field space metric satisfying the relation $\mathcal{G}^{yw} \mathcal{G}_{w x} = \delta(x-y)$. We must now find a suitable object to be the metric in the field space. To find this expression, observe that the metric and its inverse must transform as
\begin{equation}
\label{eq:fieldmettrans}
\mathcal{\widetilde G}_{xy} \ =\ K^{-1/2}_x K^{-1/2}_y\mathcal{G}_{xy}\,, \qquad 
\mathcal{\widetilde G}^{xy}\ =\ K^{1/2}_x K^{1/2}_y \mathcal{G}^{xy}\;,
\end{equation}
in order for the measure and the connection to transform correctly [cf. \eqref{measuretrans},  \eqref{conntrans}]. There is only one object in the frame-covariant formalism which transforms in this manner: the inflaton wavefunction $k(\varphi)$. Therefore, if we write
\begin{equation}
  \label{Gxy}
\mathcal{G}_{xy}\ \equiv\ k(\varphi)\, \delta(x-y)\;,
\end{equation}
then \eqref{eq:fieldmettrans} is satisfied, along with \eqref{measuretrans} and \eqref{conntrans}. With this definition, we may now find explicit expressions for the measure functional and the connection:
\begin{equation}
   \label{MGxy}
\mathcal{M}[\varphi]\ =\ \det \big(k^{1/2}(\varphi) \delta(x-y)\big)\;, \qquad 
\Gamma^z_{xy}\ =\ \frac{1}{2} \big(\ln k(\varphi)\big)_{,\varphi} \,\delta(x-y) \delta(y-z) \; ,
\end{equation}
and hence we are able to compute $\Gamma_1^{\rm VD}$ explicitly as required. 

At this point, it is important to note that the path-integral quantization of the theory from its Hamiltonian, rather its Lagrangian, gives rise to an integral measure $\mathcal{M}[\varphi]$ related to the field-space determinant of the metric $\mathcal{G}_{xy}$ given in~\eqref{Gxy}. In the same context, it is not difficult to check that  upon an arbitrary $\varphi$-reparameterization, a free theory for the field $\varphi$, where $f(\varphi ) = k(\varphi ) = 1$ and $V(\varphi ) = 0$, will still remain a free theory {\em off-shell} at the generating-functional level $\Gamma^{\rm VD}_1[\varphi, k(\varphi )]$, without inducing non-renormalizable ultra-violet infinities at the one-loop level, {\em iff} the integral measure~$\mathcal{M}[\varphi]$ as stated in~\eqref{MGxy} is chosen. Therefore, theoretical consistency of the path-integral quantization renders the Vilkovisky--DeWitt effective action {\em unique}.

Finally, let us briefly discuss the case of conformal transformations. It was shown in Section \ref{sec:modspec} that the action remains invariant under a conformal transformation [cf. \eqref{actiontransprop} with $K = 1$]. Further\-more, any functional derivative of the action with respect to $\varphi(x)$ should also remain invariant under conformal transformations, as the functional derivatives do not transform themselves. In addition, the measure functional $\mathcal{M}[\varphi]$ should not transform under conformal transformations, as there is no quantized metric perturbation in the path integral, according to the assumption~(ii) stated in the beginning of this section. Hence, under conformal transformations, we have
\begin{equation}
\Gamma_1[\tilde g_{\mu \nu}, \varphi, \tilde f(\varphi), \tilde k(\varphi), \widetilde V(\varphi)]\ =\ \Gamma_1[g_{\mu \nu}, \varphi, f(\varphi), k(\varphi), V(\varphi)]\;,
\end{equation}
which demonstrates the invariance of $\Gamma_1$ with respect to conformal transformations. 
However, the same reasoning, as outlined above, will also apply to the  one-loop Vilkovisky-DeWitt effective action $\Gamma_1^{\rm VD}$. Therefore, we conclude that $\Gamma_1^{\rm VD}$ remains invariant under frame transformations, i.e.~the combined action of both conformal transformations and inflaton reparametrizations.

\section{Conclusions}
\label{sec:disc}

We have presented a frame-covariant formalism of inflation in the slow-roll approximation for a wide class of theories known as scalar-curvature theories. We defined a set of transformations, known as {\em frame transformations}, and under these, we determined the transformation properties of the model functions: (i)~the scalar-curvature coupling function $f(\varphi)$, (ii)~the inflaton wavefunction $k(\varphi)$, and (iii)~the inflaton potential $V(\varphi)$. Consequently, we were able to show that both the classical action and its functional form remain invariant under frame transformations [cf.~\eqref{actiontransprop}], assuming that the model functions $f(\varphi)$, $k(\varphi)$ and $V(\varphi)$ transform according to \eqref{Imodelparamdef}. By generalizing the inflationary attractor solution, we have derived a new set of potential slow-roll parameters stated in~\eqref{kappaU}. Through these new parameters, we have found that inflationary observables, such as the power spectrum, the spectral indices and their runnings, can all be expressed in a concise manner in terms of the generalized potential slow-roll parameters and their $\varphi$-derivatives, which in turn depend explicitly on the model functions $f(\varphi)$, $k(\varphi)$, and $V(\varphi)$. 

In addition to obtaining concise expressions for the cosmological observables, we also utilised the potential slow-roll parameters defined in~\eqref{kappaU} to examine the effect of frame transformations on these observables in a simple manner. We have displayed that the tensor-to-scalar ratio~$r$, the spectral indices~$n_{\cal R}$ and~$n_T$, and their runnings~$\alpha_{\cal R}$ and~$\alpha_T$, are frame-invariant within this general\-ized potential slow-roll formalism, as long as the end-of-inflation condition is uniquely extended to be frame invariant as given in~\eqref{maxJF}. A direct consequence of this formalism is that one does not need to transform to the Einstein frame to utilise the potential slow-roll approximation; we have explicitly shown that this formalism reduces to the potential slow-roll approximation in the Einstein frame in Section~\ref{sec:invframe}.

To demonstrate the use of the advertised formalism, we then apply it to specific scenarios, such as the induced gravity inflation, Higgs inflation and Starobinsky-like $F(R)$ models. This application led to results for the cosmological observables which were more exact in comparison to those already presented in the literature without the need to go to the Einstein frame; our results were found to be consistent to lowest order in $1/N$, the reciprocal of the number of e-folds, with those presented in the literature.

Finally, we have outlined how our frame-covariant formalism can be naturally extended beyond the tree-level approximation within the framework of the Vilkovisky--DeWitt effective action. Specifically, we have explicitly demonstrated how the one-loop Vilkovisky--DeWitt effective action, which is written in terms of functional derivatives of the classical action, may be made invariant under inflaton reparametrizations. The Vilkovisky--DeWitt formalism is therefore the natural starting point to begin an analysis of the so-called {\it frame problem}, in addition to the study of the radiative corrections to cosmological observables and their consolidation with the slow-roll approximation. It is the authors' opinion that this is an important milestone towards the solution of the frame problem, and we hope to report progress on this issue using the Vilkovisky--DeWitt formalism in a forthcoming communication.

\subsection*{Acknowledgements}

The authors would like to thank Daniele Teresi for collaboration during the early stages of this project. The work of SK is supported by an STFC PhD studentship, and the work of AP is supported in part by the Lancaster--Manchester--Sheffield Consortium for Fundamental Physics, under STFC research grant: ST/L000520/1.

\newpage

\begin{appendix}

\section{The Inflaton Action under Frame Transformations}\label{appendix}

In this appendix, we derive the transformation properties given in~\eqref{Imodelparamdef} for the model functions~$f(\varphi )$, $k(\varphi )$ and $V(\varphi )$ describing the classical action $S$ of inflationary scalar-curvature theories [cf.~\eqref{actionJ}], under a {\it frame transformation}. As defined in Section~\ref{sec:modspec}, a {\it frame transformation} consists of a conformal transformation \eqref{weyl} and an inflaton reparameterization \eqref{Kphi}.

To start with, we first consider the classical action~\eqref{actionJ} in the Jordan frame
\begin{equation}
   \label{actionbefore}
  S[g_{\mu\nu}, \varphi, f, k,V]\ \equiv\   \int  d^4 x\,  \sqrt{-g}  \, \left[\, -\,\frac{f}{2} R\: +\: \frac{k}{2} \, g^{\mu\nu }(\nabla_\mu \varphi) (\nabla_\nu \varphi)\: -\: V\, \right]\; ,
\end{equation}
where we have suppressed the implicit dependence of the model functions $f$, $k$ and $V$ on $\varphi$. Under the conformal transformation \eqref{weyl}, $R$ transforms according to \eqref{riccitransform}.  As a consequence,  $S$ changes to
\begin{align}
  S[g_{\mu\nu}, \varphi, f, k,V]
&=\int d^4 x \, \sqrt{-g}\, \left[\, -\,\frac{f }{2} \left(\Omega^2 \widetilde R + 6\,\Omega^{-1} g^{\mu\nu} \nabla_\mu \nabla_\nu \Omega  \right)\:
+\:  \frac{k}{2}\,  g^{\mu\nu} (\partial_\mu \varphi)(\partial_\nu \varphi)\: -\: V\right]. 
\end{align}
Our next step is to rewrite $S$  in terms of $\tilde g_{\mu\nu} = \Omega^2 g_{\mu\nu}$
as follows:
\begin{eqnarray}
    \label{semitrans}
S[g_{\mu\nu}, \varphi, f, k,V]\ &=& 
  \int d^4 x \, \left\{ \sqrt{-\tilde g}\left[ \, -\, \frac{\Omega^{-2} f}{2}  \widetilde R\: +\:  \frac{k}{2}\, \Omega^{-2}\, {\tilde g}^{\mu\nu} (\partial_\mu \varphi)(\partial_\nu \varphi)\: -\: \Omega^{-4} V  \right]
\right.
\nonumber \\
& & 
\left.
 -\: 3f  \Omega^{-1} \partial_\mu (\sqrt{-g} g^{\mu\nu}  \nabla_\nu \Omega)\, 
 \vphantom{\frac{f }{2}}
 \right\}\; .
\end{eqnarray}
Then, we make use of the following identity:
\begin{equation}
  \label{identity}
  f \Omega^{-1} \partial_\mu(\sqrt{-g}\, g^{\mu\nu}\nabla_\nu \Omega) \ 
  = \
  \partial_\mu[f  \Omega^{-1} \sqrt{-g} g^{\mu\nu} \nabla_\nu \Omega]\: -\: \sqrt{-g} \,  g^{\mu\nu}\nabla_\mu [f \Omega^{-1}] \, \nabla_\nu \Omega\; .
\end{equation}
Substituting \eqref{identity} into \eqref{semitrans}, we may neglect the total derivative on the RHS of \eqref{identity}, upon total integration in the action. In addition, we assume that $\Omega$ and $\varphi$ are tempered functions that are locally Lorentz invariant and so they both depend on $x^2 \equiv x^\mu x_\mu$. Thus, the conformal factor $\Omega$ depends implicitly on the inflaton $\varphi$, i.e.~$\Omega = \Omega [\varphi (x)]$, entailing that the coordinate covariant derivative $\nabla_\mu\Omega$ can be converted into $\nabla_\mu\varphi$ through the chain rule:
\begin{equation}
   \label{eq:chain}
\nabla_\mu\Omega\ =\ \Omega_{,\varphi}\,\nabla_\mu\varphi\; .
\end{equation}
By virtue of \eqref{eq:chain}, the action $S$ in~\eqref{semitrans} becomes
\begin{eqnarray}
    \label{eq:Sinter}
S[g_{\mu\nu}, \varphi, f, k,V] &=& \int d^4 x \, \sqrt{-{\tilde g}}\: \bigg\{  
- \frac{\Omega^{-2}f}{2}\,  \widetilde R \\
&&+\:  \frac{\Omega^{-2}}{2K}\,  \Big[\, k\: +\: 6\, \Omega_{,\varphi}   
\Big(\Omega^{-1}f_{,\varphi} - f \Omega^{-2} \Omega_{,\varphi}\Big) \Big] \, 
{\tilde g}^{\mu\nu}  (\nabla_\mu\tilde\varphi)(\nabla_\nu \tilde\varphi)\: 
-\: \Omega^{-4} V\bigg\}\; ,\nonumber
\end{eqnarray}
where $\tilde \varphi = \tilde \varphi (\varphi )$ represents an arbitrary reparameterization 
of the original inflaton field $\varphi$, which is determined by the Jacobian squared:
$K = K(\varphi) = (d{\tilde\varphi}/d\varphi)^2$ [cf.\ \eqref{Kphi}].

We now observe that the last expression of the action in \eqref{eq:Sinter} can be brought into the form: 
\begin{equation}
   \label{actionafter}
S[\tilde g_{\mu\nu}, \tilde \varphi, \tilde f , \tilde k ,\widetilde V ]\
=\  \int  d^4 x\,  \sqrt{-\tilde g}  \, 
\bigg[ -\,\frac{\tilde f}{2} R \: +\: 
\frac{\tilde k}{2}  \,\tilde g^{\mu\nu }(\nabla_\mu \tilde \varphi) (\nabla_\nu \tilde \varphi ) 
\: -\: \widetilde V\,  \bigg]\; ,
\end{equation}
after making the following identifications for the transformed model functions:
\begin{align}
\tilde f (  \tilde \varphi)\ &=\ \Omega^{-2}\, f\; , \nonumber \\
\tilde k( \tilde \varphi)\ &=\ \frac{\Omega^{-2}}{K}\, \Big( k \:
-\: 6\, f\,\Omega^{-2}\Omega_{,\varphi}^2\:
+\:  6\, \Omega^{-1}    f_{,\varphi}\, \Omega_{,\varphi}  \Big)\;,\tag{\ref{Imodelparamdef}}\\
\widetilde V(  \tilde\varphi)\ &=\ \Omega^{-4}\, V\; .\nonumber
\end{align}
Notice that the tilted model functions are evaluated at $\tilde \varphi$, which is achieved by expressing $\varphi$ as $\varphi = \varphi (\tilde \varphi )$, e.g.
\begin{equation}
\tilde f (\tilde \varphi )\ =\ \Omega^{-2} [\varphi (\tilde \varphi )]\: f[\varphi (\tilde \varphi )]\; .
\end{equation}
Evidently, the latter ensures that the actions \eqref{actionbefore} and \eqref{actionafter} are equal, exhibiting the same functional dependence~[cf.\ \eqref{actiontransprop}]. We may now specialize the frame transformations~\eqref{Imodelparamdef}, so as to go to the Einstein frame. This is accomplished by choosing $\Omega^2 = f$, such that $\tilde f = 1$, and the squared Jacobian $K$, such that the inflaton kinetic term becomes canonical, with $\tilde k = 1$.

\end{appendix}

\end{document}